\documentclass{aa}
\usepackage{graphicx}
\usepackage{txfonts}
\usepackage{longtable}
\usepackage{amssymb}
\usepackage{natbib}
\begin{document}
   \title{SN 2009E: a faint clone of SN 1987A}

   \author{A. Pastorello\inst{1,2,3}\thanks{email: andrea.pastorello@oapd.inaf.it}
           \and M. L. Pumo\inst{1,4}
	   \and H. Navasardyan\inst{1}
           \and L. Zampieri\inst{1}
	   \and M. Turatto\inst{5}
	   \and J. Sollerman\inst{6}
	   \and F. Taddia\inst{6}
	   \and E. Kankare\inst{7}
           \and S. Mattila\inst{7}
           \and J. Nicolas\inst{8}
	   \and E. Prosperi\inst{9}
           \and A. San Segundo Delgado\inst{10}
	   \and S. Taubenberger\inst{11}
	   \and T. Boles\inst{12}
           \and M. Bachini\inst{13}
	   \and S. Benetti\inst{1}
	   \and F. Bufano\inst{4} 
           \and E. Cappellaro\inst{1} 
	   \and A. D. Cason\inst{14}
           \and G. Cetrulo\inst{15}
	   \and M. Ergon\inst{6}
	   \and L. Germany\inst{16}
           \and A. Harutyunyan\inst{17}
	   \and S. Howerton\inst{18}
	   \and G. M. Hurst\inst{19}
	   \and F. Patat\inst{20}
	   \and M. Stritzinger\inst{6,21}
	   \and L.-G. Strolger\inst{22}
	   \and W. Wells\inst{23}
}
   \offprints{A. Pastorello}

   \institute{INAF - Osservatorio Astronomico di Padova, Vicolo dell' Osservatorio 5, I-35122, Padova, Italy
   \and
   Astrophysics Research Centre, School of Mathematics and Physics, Queen's University Belfast, Belfast BT7 1NN, United Kingdom
   \and  
   Dipartimento di Astronomia, Universit\`a di Padova, Vicolo dell'Osservatorio 3, I-35122 Padova, Italy       
   \and         
   INAF - Osservatorio Astrofisico di Catania, via S. Sofia 78, I-95123 Catania, Italy
   \and   
   INAF - Osservatorio Astronomico di Trieste, Via G. B. Tiepolo 11, I-34143, Trieste, Italy 
   \and
   Oskar Klein Centre, Department of Astronomy, AlbaNova, Stockholm University, SE-10691, Stockholm, Sweden
   \and
   Tuorla Observatory, Department of Physics \& Astronomy, University of Turku, V\"ais\"al\"antie 20, FI-21500, Piikki\"o, Finland
   \and
   Les Mauruches Observatoire, 364 Chemin de Notre Dame, F-06220, Vallauris, France 
   \and
   Osservatorio Astronomico di Castelmartini, IAU 160, Via Bartolini 1317, I-51036, Larciano, Pistoia, Italy
   \and
   Observatorio El Gujio, Onice 21, E-28260, Galapagar, Madrid, Spain
   \and
   Max-Planck-Institut f\"ur Astrophysik, Karl-Schwarzschild-Str. 1, D-85741, Garching bei M\"unchen, Germany
   \and
   Coddenham Astronomical Observatory, Suffolk, United Kingdom
   \and
   Osservatorio Astronomico di Tavolaia, Piazza della Vittoria 41, I-56020 Santa Maria a Monte, Pisa, Italy
   \and
   private address, 105 Glen Pine Trail, Dawnsonville, GA 30543, USA
   \and
   Osservatorio Astronomico Polse di Cougnes, Zuglio, I-33020 Udine, Italy
   \and  
   Center for Astrophysics and Supercomputing, Swinburne University of Technology, Hawthorn, VIC 3122, Australia 
   \and
   Fundaci\'on Galileo Galilei - INAF, Telescopio Nazionale Galileo, 38700 Santa Cruz de la Palma, Tenerife, Spain
   \and
   private address, 1401 South A, Arkansas City, KS 67005, USA 
   \and
   The Astronomer, 16 Westminster Close, Basingstoke, Hants, RG22 4PP, United Kingdom
   \and
   European Southern Observatory, Karl-Schwarzschild-Str. 2, D-85748, Garching bei M\"unchen, Germany
   \and 
   Dark Cosmology Centre, Niels Bohr Institute, University of Copenhagen, Juliane Maries Vej 30, 2100 Copenhagen, Denmark
   \and
   Department of Physics and Astronomy, Western Kentucky University, 1906 College Heights Blvd., Bowling Green, KY 42101-1077, USA 
   \and
   University of Oklahoma, Health Science Center, 1100 N. Lindsay, Oklahoma City, OK 73104, USA}

   \date{Received  XXXXX ; accepted XXXXX}

 
  \abstract
   {1987A-like events form a rare sub-group of hydrogen-rich core-collapse supernovae that are thought
to originate from the explosion of blue supergiant stars. Although SN 1987A is the best known
supernova, very few objects of this group have been discovered and, hence, studied.} 
   {In this paper we investigate the properties of SN 2009E, which exploded in a relatively nearby spiral
galaxy (NGC 4141) and that is probably the faintest
1987A-like supernova discovered so far. 
We also attempt to characterize this subgroup of core-collapse supernovae with the help of the literature 
and present new data for a few additional objects.}
   {The lack of early-time observations from professional telescopes is compensated by
frequent follow-up observations performed by a number of amateur astronomers. This allows us to reconstruct
a well-sampled light curve for SN 2009E. Spectroscopic observations which started about 2 months
after the supernova explosion, highlight significant differences
between SN 2009E and the prototypical SN 1987A. Modelling the data of SN 2009E allows us to constrain the explosion parameters
and the properties of the progenitor star, and compare the inferred estimates with those available for the similar SNe 1987A and 1998A.}
   {The light curve of SN 2009E is less luminous than that of SN 1987A and the other members of this
class, and the maximum light curve peak is reached at a slightly later epoch than in SN 1987A. Late-time
photometric observations suggest that SN 2009E ejected about 0.04 M$_\odot$ of $^{56}$Ni,
which is the smallest $^{56}$Ni mass in our sample of 1987A-like events.
Modelling the observations with a radiation hydrodynamics code, we infer
for SN 2009E a kinetic plus thermal energy of about 0.6 foe, an initial radius of $\sim 7 \times 10^{12}$ cm and an ejected mass
of $\sim$ 19 M$_\odot$.
The photospheric spectra show a number of narrow (v$\approx$1800 km s$^{-1}$) metal lines, with unusually
strong Ba II lines. The nebular spectrum displays narrow emission lines of H, Na I, [Ca II] and [O I],
with the [O I] feature being relatively strong compared to the [Ca II] doublet. 
The overall spectroscopic evolution is reminiscent of that of the faint $^{56}$Ni-poor type II-plateau supernovae. 
This suggests that SN 2009E belongs to the low-luminosity, low $^{56}$Ni mass, low-energy tail in the 
distribution of the 1987A-like objects in the 
same manner as SN 1997D and similar events represent the faint tail in the distribution of physical properties 
for normal type II-plateau supernovae. }



\maketitle
%

\section{Introduction}

The explosion of supernova (SN) 1987A in the Large Magellanic Cloud (LMC)
was an epic event not only because it was the nearest SN in
a period of about 4 centuries, but also because it changed
significantly the general understanding of the destiny of massive stars.
It was commonly believed that hydrogen-rich (H-rich) type II plateau
supernovae (SNe IIP) were generated by the explosion of red supergiant
(RSG) stars. However, the unusual photometric evolution of SN 1987A, with
a broad light curve peak reached about 3 months after the explosion instead of the
classical plateau observed in H-rich SNe, and the direct detection
of the progenitor star in pre-explosion images changed this general view
and proved that the precursor of this SN in the LMC was instead a blue
supergiant (BSG, Arnett et al. 1989 and references therein).

Due to their intrinsic rarity and faint early time luminosity,
only very few 1987A-like objects were  discovered in the past.
Good datasets therefore exist only for a handful of objects
(SN 1998A, Woodings et al. 1998, Pastorello et al. 2005; SN 2000cb, Hamuy
2001, Kleiser et al. 2011; SNe 2006V and 2006au, Taddia et al. 2011). 
In analogy with SN 1987A, some of these 
objects show broad, delayed light curve peaks in all bands (e.g. SN 1998A), 
while others (e.g. SN 2000cb and SN 1982F) show 1987A-likeness only in the red 
bands, while the blue bands display a rather normal type IIP behaviour.

In this context, the discovery of a new sub-luminous SN with a light curve
comparable to that of SN 1987A, but with other observed properties (fainter
intrinsic luminosity at peak,  smaller synthesized  $^{56}$Ni mass
and lower expansion velocity of the ejected material) resembling
those of some underluminous type IIP SNe (Turatto et al. 1998, Benetti et
al. 2001, Pastorello et al. 2004, 2006, 2009) is even more interesting.

This paper is organized as follows: in Section \ref{obs} we give basic information on 
SN 2009E and its host galaxy, and we present photometric and spectroscopic observations
of the SN. In Section \ref{const_pro} we discuss the properties of the progenitor star and 
the explosion parameters as derived from the characteristics of the nebular spectrum and by modelling the SN observations. In 
Section \ref{discussion} we analyse the general properties of the family of SNe similar to 
SN 1987A, whilst the rate of these events is computed in Section \ref{rate}.  Main conclusions are given in Section \ref{summary}.
Finally, an Appendix has been included to present our sample of 1987A-like events
(Appendix \ref{family}), and to compare the light curves of well monitored objects with those
of SN 1987A (Appendix \ref{appB}).

\section{Observations of SN 2009E} \label{obs}

   \begin{figure}
   \centering
   \includegraphics[angle=0,width=8.8cm]{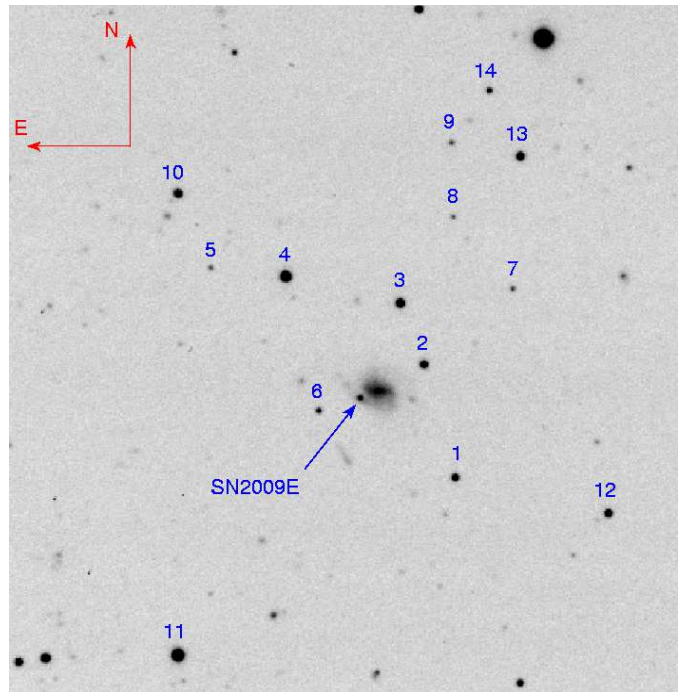}
   \caption{SN 2009E in NGC 4141. Unfiltered image obtained on April 29, 2009 by J. N. 
   with a 0.28-m f/6.5 reflector equipped with a ST8XME with Kaf1602E CCD. Our  
  sequence of reference stars is marked by numbers. The field of view is about $10' \times 10'$.}
              \label{Fig1}
    \end{figure}

\subsection{The host galaxy} \label{hg}

NGC 4141 is classified by HyperLeda\footnote{{\it http://leda.univ-lyon1.fr/}} as an SBc galaxy,  
rich in H II regions (NED\footnote{{\it http://nedwww.ipac.caltech.edu/index.html}}). 
\citet{kew05} estimated the integrated oxygen abundance to be log(O/H) + 12 = 8.60 (8.74 in the nucleus).
They also estimated a star formation rate (SFR) from the integrated flux of H$\alpha$, viz. $\sim$ 0.6 M$_\odot$ yr$^{-1}$.

Adopting the recessional velocity corrected for
Local Group infall into Virgo quoted by HyperLeda (v$_{Vir}$ = 2158 $\pm$ 20 km s$^{-1}$) and a Hubble
constant H$_0$ = 72 $\pm$ 5 km s$^{-1}$ Mpc$^{-1}$, we obtain a distance of about 29.97 $\pm$ 2.10 Mpc 
(i.e. distance modulus $\mu$ = 32.38 $\pm$ 0.35 mag). 
The Galactic extinction in the direction of NGC 4141 is very low, i.e. $E(B-V)$ = 0.02 mag \citep{sch98}.
Narrow interstellar Na I $\lambda\lambda$5889,5895 (hereafter Na ID) absorption at the host galaxy rest 
wavelength is marginally detected in the SN spectra, with an equivalent width (EW) of about 0.12 \AA. 
Adopting the relation  between EW$_{Na ID}$ and interstellar extinction  from  \cite{tura03}, we find
a host galaxy reddening of $E(B-V)$ = 0.02 mag\footnote{Note, however, that \citet{poz11}
casted doubt on the robustness of the correlation between EW (Na ID) and interstellar extinction as determined from low resolution 
type Ia SN spectra.}. The total colour excess in the direction of SN 2009E
is therefore $E(B-V)$ = 0.04 mag. 

\subsection{The discovery of SN 2009E} \label{sn09e}

SN 2009E was discovered in the spiral galaxy NGC 4141 on January 3rd, 2009,
 at an unfiltered magnitude of 17.8 \citep{bol09}. With the distance and reddening estimated in Section \ref{hg},
the absolute magnitude at the discovery was about $-$14.7.
The position of SN 2009E is $\alpha = 12^h09^m49^s.56 \pm 0^s.03$, $\delta = +58^\circ50'50\farcs3 \pm 0\farcs1$ (equinox J2000.0),
which is $17\farcs2$ East and $6\farcs9$ South of the center of the host galaxy.
According to \citet{bol09}, nothing was visible in pre-explosion images obtained on 2008 February 6th (to limiting magnitude of 19.5).
We also found pre-explosion archive $g'$, $R_c$ and $I_c$  images\footnote{Images have been downloaded through the SMOKA Data Archive \protect\citep{baba02}.} obtained on 2008 February 24th at the 0.5-m telescope of the 
Akeno Observatory/ICRR (Yamanashi, Japan) and no source brighter 
than $V$ = 19.5, $R$ = 19.6, $I$ = 18.8 mag was visible at the SN position. In this paper, we will adopt January 1.0 UT ($JD$=2454832.5, see Section \ref{lc}) as an indicative epoch for the core-collapse.

NGC 4141 also hosted the type II SN 2008X 
\citep{bol08,mad08,blo08}, that exploded $7\farcs6$ East and $4\farcs6$ North of the nucleus of the galaxy.
Interestingly, SN 2008X was another
sub-luminous \citep[$M_R \approx -$14.9,][]{bol08} event. The 
classification spectrum of SN 2008X obtained a few weeks after the explosion
showed that it was a type II SN \citep{blo08}, although with rather narrow spectral lines, 
similar to those observed in SN 2005cs \citep{pasto06}. 

Probably the faint apparent magnitude of SN 2009E discouraged astronomers to promptly classify this
object. However, later on, \citet{pro09} noted that SN 2009E brightened
by 1$-$1.5 mag during the subsequent two months, and only at about 80 days
from the discovery the object was classified by \cite{nava09} as a type
II SN around the end of the H recombination phase. SN 2009E appeared to show some
similarity with the peculiar SN 1987A, especially in the late-time light curve
brightening and the unusual strength of the Ba II P-Cygni lines. The
most obvious differences were in the lower luminosity and the much narrower
spectral features.

\subsection{Light Curves} \label{lc}

Whilst professional astronomers initially missed to follow this transient object,
amateur astronomers monitored SN 2009E extensively, and it is thanks to their efforts
that we can now recover information on the early flux evolution.
Most images collected during the period
January 2009 to May 2009 were from amateur observations. Since the end of March, we started multiband
photometric and spectroscopic follow-up observations with larger,
professional telescopes. All photometric data  were reduced with standard
techniques in IRAF\footnote{IRAF is distributed by the National Optical Astronomy Observatories,
which are operated by the Association of Universities for Research
in Astronomy, Inc., under cooperative agreement with the National
Science Foundation.}. Images were first overscan, bias and flat-field 
corrected. Then, photometric measurements of filtered
images were performed using a PSF-fitting technique because suitable 
template images were not available. However, since the SN location was quite peripheral 
in the galaxy arm, its luminosity largely exceeded that of the surroundings, except at 
very late epochs when the SN became weak. We estimated the 
uncertainty due to the non-flat background by placing in all images several artificial 
stars close to the SN location. 
The adopted errors were the r.m.s. of the recovered artificial star PSF magnitudes.
Multi-band zero-points for the different nights were computed through observations of
 standard stellar fields \citep{lan92} obtained during the same nights as the SN
observations. The SN magnitudes were finally fine tuned with reference to 
a sequence of stars in the field of NGC 4141 (see
Figure 1), calibrated by averaging magnitudes measured in selected photometric nights.
The magnitudes of the local sequence stars are shown in Table \ref{Tab1},
while the final SN magnitudes are reported in Table \ref{Tab2}.

 \begin{figure}
   \centering
   \includegraphics[angle=0,width=9.6cm]{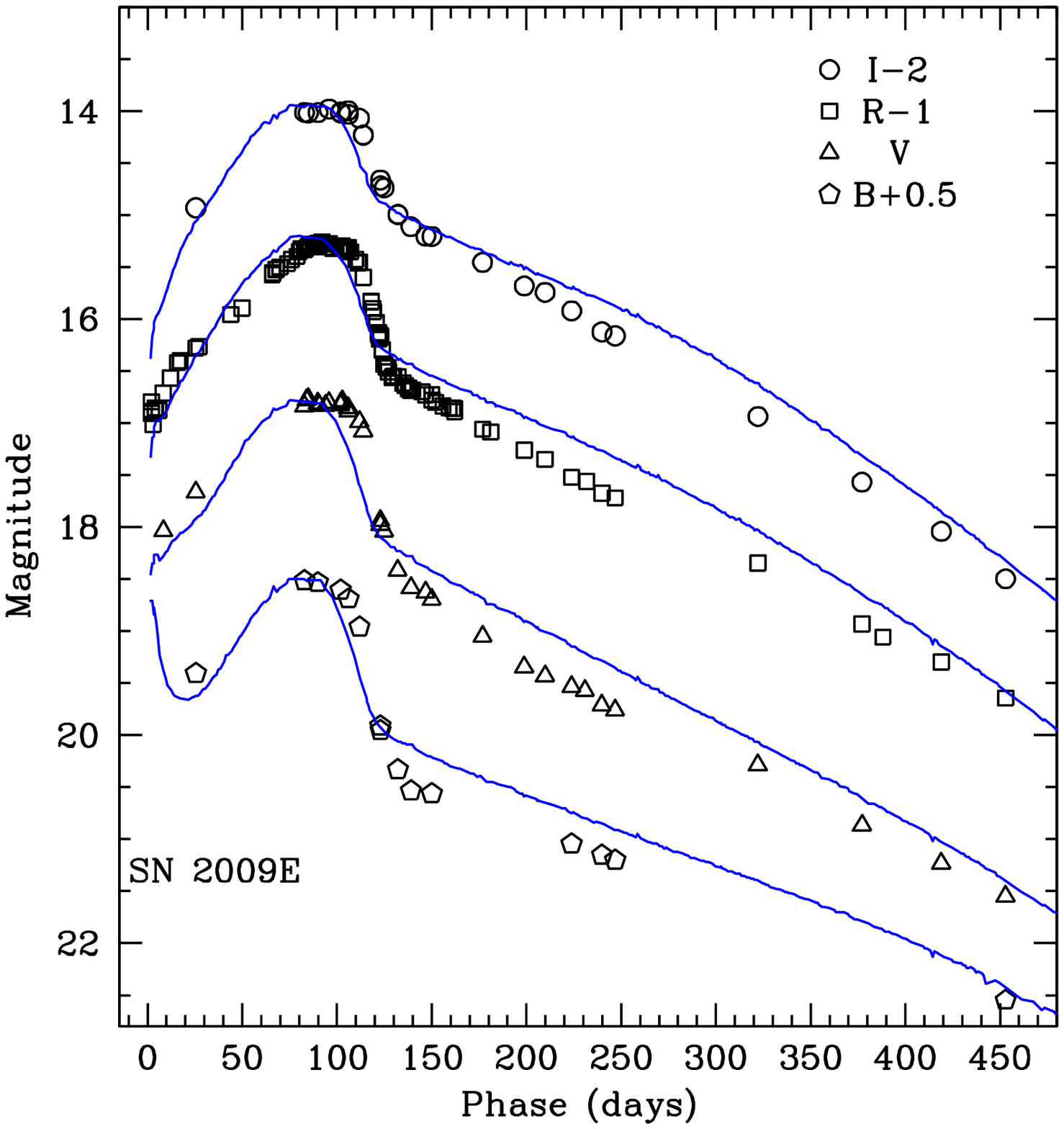}
   \caption{$B$, $V$, $R$ and $I$-band light curves of SN 2009E compared with those of 
   SN 1987A \protect\citep[solid blue line;][]{men87,cat87,cat88,cat89,whi88,whi89}, shifted arbitrarily 
   in magnitude to match the peak magnitudes of SN 2009E.}
              \label{Fig2}
\end{figure}

 \begin{figure}
   \centering
   \includegraphics[angle=0,width=9.2cm]{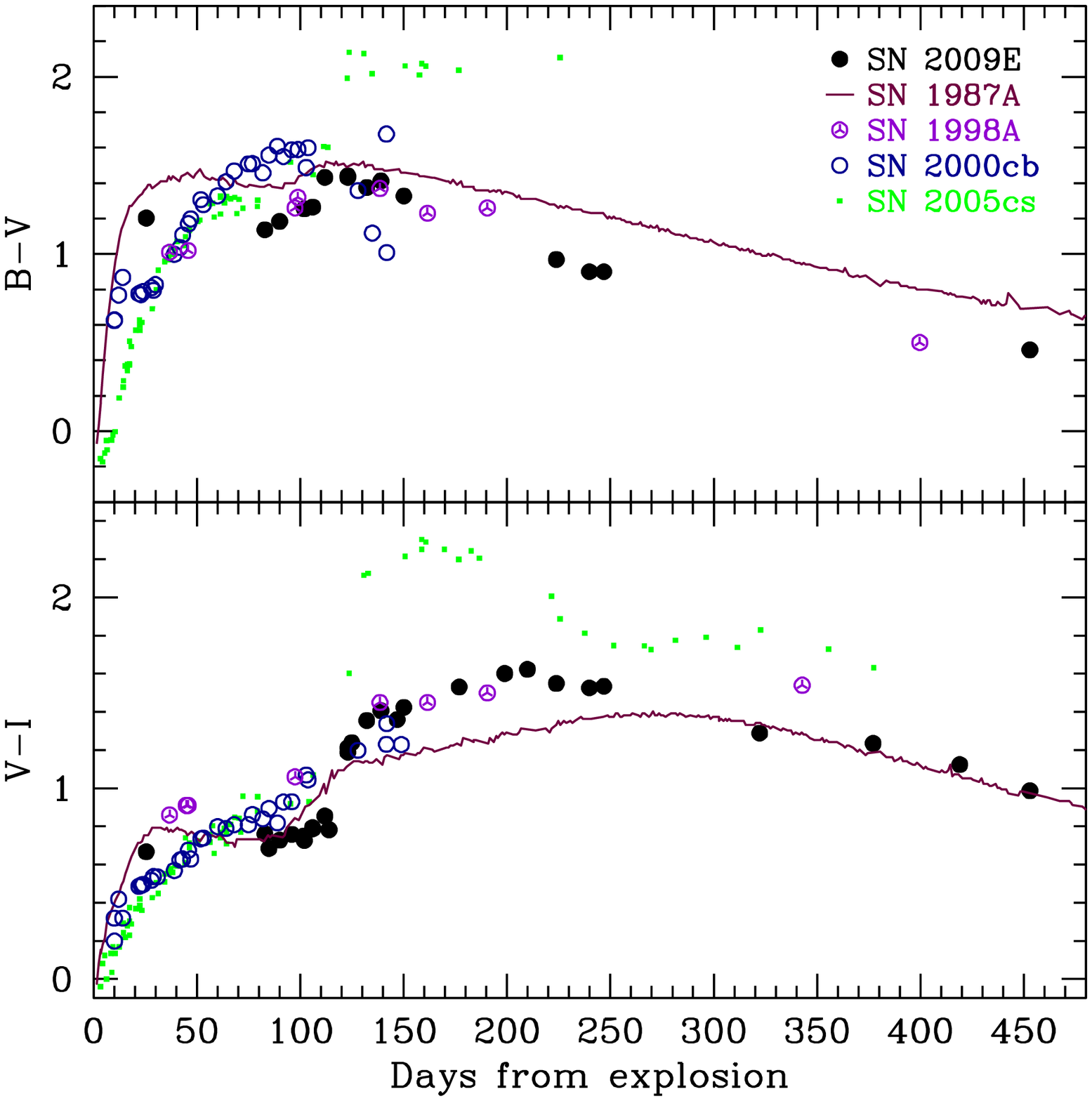}
   \caption{$B-V$ (top) and $V-I$ (bottom) colour curves of SN 2009E compared with those of a
   subsample of SN~1987A-like events with multi-colour photometric coverage: SNe 1987A \protect\citep{men87,cat87,cat88,cat89,whi88,whi89}, 1998A \protect\citep{pasto05}, and 2000cb \protect\citep{ham01}.
   The colour curves of the sub-luminous type IIP SN 2005cs \protect\citep{tsv06,pasto09} is also shown for comparison.}
              \label{Fig3}
\end{figure}

\begin{table*}
\centering
\caption{Magnitudes of reference stars in the SN field, calibrated on several photometric nights. The root mean square of the average magnitudes are reported in brackets.}\label{Tab1}
\begin{tabular}{cccccc} \hline\hline
 Star & $U$ & $B$ & $V$ & $R$ & $I$  \\ \hline
1 & 18.86 (0.03) & 17.59 (0.02) & 16.52 (0.02) & 15.74 (0.01) & 15.18 (0.01) \\  
2 & 17.47 (0.02) & 16.88 (0.01) & 16.11 (0.01) & 15.60 (0.01) & 15.20 (0.01) \\
3 & 15.85 (0.04) & 15.87 (0.01) & 15.35 (0.01) & 15.01 (0.01) & 14.69 (0.01) \\
4 & 18.07 (0.03) & 16.89 (0.01) & 15.41 (0.02) &   --         & -- \\
5 & 18.81 (0.03) & 18.73 (0.01) & 18.05 (0.02) & 17.62 (0.01) & 17.19 (0.02) \\ 
6 & 17.75 (0.03) & 17.85 (0.01) & 17.26 (0.01) & 16.83 (0.01) & 16.43 (0.01) \\
7 & 20.76 (0.05) & 19.33 (0.02) & 18.17 (0.01) & 17.31 (0.01) & 16.56 (0.02) \\
8 & 18.97 (0.05) & 18.91 (0.02) & 18.23 (0.01) & 17.78 (0.01) & 17.37 (0.01) \\
9 &     --           & 19.90 (0.02) & 18.46 (0.02) & 17.47 (0.01) & 16.58 (0.02) \\
10 &    --           & 16.32 (0.01) & 15.70 (0.02) & 15.31 (0.02) & 14.94 (0.01) \\
11 &    --           & 14.45 (0.02) & 13.73 (0.03) & 13.15 (0.04) & 12.83 (0.01) \\
12 &    --           & 16.43 (0.03) & 15.86 (0.03) & 15.48 (0.01) & 15.11 (0.02) \\ 
13 & 16.63 (0.06)  & 16.40 (0.01) & 15.67 (0.02) & 15.22 (0.01) & 14.78 (0.02) \\
14 &    --           & 19.86 (0.06) & 18.27 (0.10) &    --            &     --           \\ 
\hline
\end{tabular}
\end{table*}

   \begin{figure*}
   \centering
   \includegraphics[angle=270,width=14.0cm]{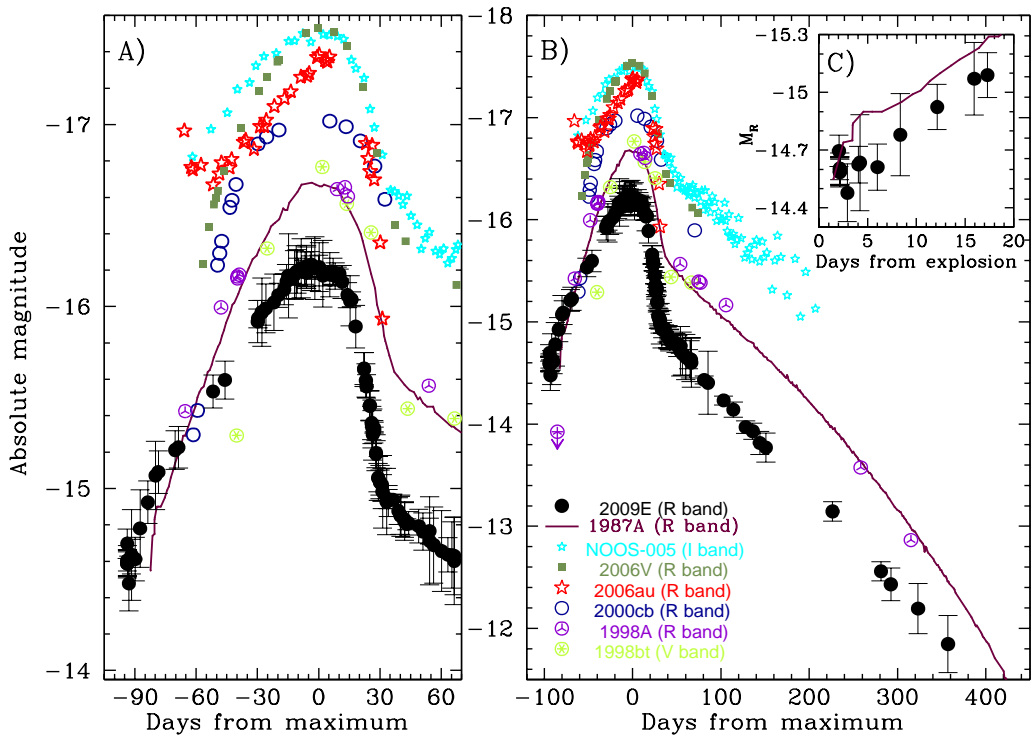}
   \caption{$R$-band absolute light curve of SN 2009E compared with the absolute light curves of a
   number of SN~1987A-like events: NOOS-005 ($I$ band, OGLE collaboration), SNe 2006V, 2006au, 1998bt
   \citep[$V$ band, L. Germany, private communication, see also][]{ger98}, 2000cb, 1998A
  and the prototype 1987A.
   Panel A: detail on the broad light curve peak. Panel B: the full light curve evolution. Panel C: 
   blow-up of the very early-time light curves of SNe 2009E and 1987A, soon after shock break-out.}
              \label{Fig4}
    \end{figure*}

Since a number of unfiltered pre-explosion images of NGC 4141 were available,
magnitude measurements on the amateurs' images were performed with the PSF-fitting technique 
after subtraction of the best seeing host galaxy template image. In this way we minimized the
contamination of the background sources near the SN location. The subsequent
photometric calibration was performed with the prescriptions
discussed in \citet{pasto08}. Since the detectors used by amateur
astronomers in their follow-up campaign of SN 2009E have quantum efficiencies
preferentially peaking at $\sim$ 5700$-$6500 \AA, unfiltered magnitudes were scaled to
match the $R$-band photometry. 
The SN magnitude errors accounted for the uncertainties in this convertion.
The final $R$-band-calibrated magnitudes are reported in Table \ref{Tab2},
marked with an asterisk. The $B$, $V$, $R$, $I$ light curves
of SN 2009E, compared with those of SN 1987A, are shown in Figure \ref{Fig2}.

In Figure \ref{Fig3} we compare the $B-V$ and $V-I$ colour curves of SN 2009E with those of a subsample of
1987A-like events and with the sub-luminous type IIP SN 2005cs.
The colour curves of SN 2009E are remarkably similar to those of SN 1987A at all phases. The colour estimates at phase $\sim$ 25 days  ($B-V \approx$ 1.2 and $V-I \approx$ 0.7 mag)
suggest that, in analogy with SN 1987A, SN 2009E became red in a time scale that is shorter than for classical SNe IIP 
(about 1 month for SN 2009E, while  $\sim$2 months for type IIP SNe).
Between day 30 and day 110, during the recombination phase, the colours remain almost constant. Then, with the post-peak luminosity decline, the SN becomes redder again, reaching $B-V \approx$ 1.4 between 110 and 140 days after explosion (this is remarkably similar to the4 $B-V$ colour observed in SN 1987A). 
The $V-I$ colour, instead, peaks at about 200 days ($V-I \approx$ 1.6 mag), which is a bit earlier than observed for SN 1987A ($V-I \approx$ 1.4 mag between 230 and 310 days).
Subsequently, the colours turn back to the blue again.

\newpage

\longtabL{2}{
\setlength{\LTcapwidth}{7in}
\setcounter{LTchunksize}{59}
\begin{longtable}{cccccccc}
\caption{ Calibrated $U$, $B$, $V$, $R$, $I$ band magnitudes of SN 2009E. 
The numbers in column 8 identify the different instrumental configurations. Unfiltered observations rescaled to
R-band magnitudes are  marked with
the symbol ``$\star$''. The ``$\star\star$'' symbol marks two epochs in which both filtered (V or I band) and 
unfiltered observations were collected.
The symbol ``$\ddag$'' marks a $g'$-band observation converted into Johnson-Bessell $V$ band.}\label{Tab2} \\ \hline\hline
Date & $JD$  & $U$ & $B$ & $V$ & $R$ & $I$ & Instrument \\ \hline
Feb06 2008 & 2454502.5  &  -- &  -- &  -- &  $>$19.5     &  -- & 1$\star$ \\
Feb24 2008 & 2454521.22 & -- & -- & $>$19.5$\ddag$ & $>$19.6 & $>$18.9 & 2 \\
Jan03 2009  & 2454834.56 &  -- &  -- &  -- &  17.79  (0.08)&  -- &  1$\star$ \\
Jan03 2009  & 2454834.63 &  -- &  -- &  -- &  17.89  (0.06)&  -- &  1$\star$ \\
Jan03 2009  & 2454834.63 &  -- &  -- &  -- &  17.89  (0.08)&  -- &  1$\star$ \\
Jan03 2009  & 2454834.64 &  -- &  -- &  -- &  17.90  (0.08)&  -- &  1$\star$ \\
Jan03 2009  & 2454834.64 &  -- &  -- &  -- &  17.91  (0.07)&  -- &  1$\star$ \\
Jan03 2009  & 2454834.65 &  -- &  -- &  -- &  17.90  (0.08)&  -- &  1$\star$ \\
Jan03 2009  & 2454835.42 &  -- &  -- &  -- &  18.01  (0.15)&  -- &  1$\star$ \\
Jan05 2009  & 2454836.56 &  -- &  -- &  -- &  17.87  (0.10)&  -- &  3$\star$ \\
Jan05 2009  & 2454836.71 &  -- &  -- &  -- &  17.86  (0.25)&  -- &  4$\star$ \\ 
Jan07 2009  & 2454838.51 &  -- &  -- &  -- &  17.88  (0.12)&  -- &  5$\star$ \\ 
Jan09 2009  & 2454840.84 &  -- &  -- & 18.04 (0.33)&  17.71  (0.21)&  -- & 6 \\
Jan13 2009  & 2454844.63 &  -- &  -- &  -- &  17.57  (0.12)&  -- &  4$\star$ \\
Jan16 2009  & 2454848.44 &  -- &  -- &  -- &  17.42  (0.19)&  -- &  6$\star$ \\
Jan18 2009  & 2454849.80 &  -- &  -- &  -- &  17.40  (0.12)&  -- &  7$\star$ \\
Jan26 2009  & 2454858.01 & $>$20.35 &18.91 (0.29)& 17.66 (0.13)&  17.28  (0.11)& 16.93 (0.07)& 8 \\
Jan28 2009  & 2454859.64 &  -- &  -- &  -- &  17.26  (0.11)&  -- &  9$\star$ \\
Feb14 2009  & 2454876.56 &  -- &  -- &  -- &  16.96  (0.09)&  -- &  10$\star$\\
Feb19 2009  & 2454882.39 &  -- &  -- &  -- &  16.89  (0.10)&  -- &  11$\star$ \\
Mar07 2009  & 2454898.48 &  -- &  -- &  -- &  16.57  (0.12)&  -- &  10$\star$ \\
Mar08 2009  & 2454898.55 &  -- &  -- &  -- &  16.56  (0.05)&  -- &  6$\star$ \\
Mar09 2009  & 2454900.42 &  -- &  -- &  -- &  16.52  (0.07)&  -- &  6$\star$\\
Mar09 2009  & 2454900.46 &  -- &  -- &  -- &  16.52  (0.19)&  -- &  12$\star$ \\
Mar11 2009  & 2454902.50 &  -- &  -- &  -- &  16.50  (0.13)&  -- &  10$\star$ \\
Mar15 2009  & 2454906.44 &  -- &  -- &  -- &  16.47  (0.16)&  -- &  10$\star$ \\
Mar18 2009  & 2454908.58 &  -- &  -- &  -- &  16.43  (0.07)&  -- &  13$\star$ \\
Mar20 2009  & 2454911.49 &  -- &  -- &  -- &  16.40  (0.08)&  -- &  10$\star$\\
Mar22 2009  & 2454912.57 &  -- &  -- &  -- &  16.35  (0.09)&  -- &  9$\star$ \\
Mar23 2009  & 2454913.50 &  -- &  -- &  -- &  16.32  (0.23)&  -- &  14$\star$ \\
Mar23 2009  & 2454913.67 &  -- &  -- &  -- &  16.33  (0.04)&  -- &  15$\star$ \\
Mar24 2009  & 2454915.41 &  -- &18.02 (0.10)& 16.84 (0.03)&  16.33  (0.03)& 16.01 (0.02)& 16 \\
Mar25 2009  & 2454916.42 &  -- &  -- &  -- &  16.32  (0.13)&  -- &  10$\star$ \\
Mar25 2009  & 2454916.46 &  -- &  -- & 16.78 (0.05)&   -- &  -- &  17 \\
Mar26 2009  & 2454917.33 &  -- &  -- & 16.77 (0.14)&  16.30  (0.15)& 16.02 (0.17)&  17$\star$ \\
Mar27 2009  & 2454917.73 &  -- &  -- &  -- &  16.30  (0.12)&  -- &  15$\star$ \\
Mar27 2009  & 2454918.38 &  -- &  -- &  -- &  16.30  (0.07)&  -- &  3$\star$ \\
Mar29 2009  & 2454919.75 &  -- &  -- &  -- &  16.28  (0.16)&  -- &  15$\star$\\
Mar30 2009  & 2454921.42 &  -- &  -- &  -- &  16.30  (0.07)&  -- &  3$\star$ \\
Mar31 2009  & 2454921.70 &  -- &  -- &  -- &  16.29  (0.10)&  -- &  15$\star$ \\
Mar31 2009  & 2454922.36 &  -- &  -- &  -- &  16.27  (0.12)&  -- &  17$\star$\\
Mar31 2009  & 2454922.43 &  -- &  -- &  -- &  16.27  (0.11)&  -- &  10$\star$\\ 
Mar31 2009  & 2454922.49 &  -- &18.03 (0.02)& 16.81 (0.01)&  16.31  (0.01)& 16.02 (0.01)&  18 \\
Apr01 2009  & 2454923.34 &  -- &  -- & 16.83 (0.04)&  16.29  (0.04)&  -- &  19 \\
Apr02 2009  & 2454924.39 &  -- &  -- &  -- &  16.27  (0.06)&  -- &  3$\star$ \\
Apr02 2009  & 2454924.44 &  -- &  -- &  -- &  16.26  (0.07)&  -- &  17$\star$ \\
Apr03 2009  & 2454924.75 &  -- &  -- &  -- &  16.26  (0.14)&  -- &  15$\star$ \\ 
Apr04 2009  & 2454926.42 &  -- &  -- & 16.83 (0.08)&   -- &  -- &  19 \\ 
Apr04 2009  & 2454926.45 &  -- &  -- &  -- &  16.28  (0.17)&  -- &  10$\star$\\ 
Apr06 2009  & 2454928.33 &  -- &  -- & 16.80 (0.06)&   -- & 15.98 (0.03)&  19 \\  
Apr06 2009  & 2454928.34 &  -- &  -- &  -- &  16.28  (0.07)&  -- &  3$\star$ \\ 
Apr07 2009  & 2454928.61 &  -- &  -- &  -- &  16.27  (0.13)&  -- &  9$\star$\\ 
Apr08 2009  & 2454930.32 &  -- &  -- &  -- &  16.31  (0.18)&  -- &  17$\star$ \\ 
Apr08 2009  & 2454930.42 &  -- &  -- &  -- &  16.32  (0.06)&  -- &  10$\star$ \\  
Apr12 2009  & 2454933.79 &  -- &  -- &  -- &  16.32  (0.12)&  -- &  15$\star$ \\ 
Apr12 2009  & 2454934.39 &  -- &  -- & 16.82 (0.04)&   -- & 16.01 (0.03)&  19 \\  \hline
\\
\caption{continued.}\\ \hline\hline
Date & $JD$ & $U$ & $B$ & $V$ & $R$ & $I$ & Instrument \\ \hline
Apr12 2009  & 2454934.46 &19.53 (0.11) &18.10 (0.02)& 16.81 (0.01)&  16.300 (0.01) & 16.02 (0.02)&  18 \\
Apr13  2009      & 2454935.36 &  -- &  -- & 16.79 (0.07)&  16.300 (0.08)&  -- & 6$\star\star$ \\ 
Apr13  2009      & 2454935.40 &  -- &  -- &  -- &  16.32  (0.05)&  -- &  3$\star$ \\ 
Apr14  2009      & 2454936.37 &  -- &  -- &  -- &  16.32  (0.07)&  -- &  17$\star$ \\
Apr15  2009      & 2454937.34 &  -- &  -- &  -- &  16.33  (0.05)&  -- &  17$\star$ \\
Apr16  2009      & 2454938.41 &  -- &  -- & 16.85 (0.03)&  16.31  (0.03)& 16.00 (0.02)&  20 \\ 
Apr17  2009      & 2454938.53 &  -- &18.19 (0.02)& 16.88  (0.01)&  16.33 (0.01)& 16.03 (0.01)&  18 \\ 
Apr18  2009      & 2454939.76 &  -- &  -- &  -- &  16.36  (0.09)&  -- &  15$\star$ \\
Apr20  2009      & 2454942.39 &  -- &  -- &  -- &  16.43  (0.07)&  -- &  3$\star$ \\ 
Apr22  2009      & 2454943.74 &  -- &  -- &  -- &  16.46  (0.14)&  -- &  15$\star$ \\
Apr22  2009      & 2454944.32 &  -- &  -- &  -- &  16.45  (0.14)&  -- &  6$\star$ \\ 
Apr22  2009      & 2454944.38 &  -- &18.46 (0.35)& 16.99 (0.10)&  16.45  (0.04)& 16.07 (0.08)&  6 \\
Apr24  2009      & 2454946.40 &  -- &  -- & 17.08 (0.07)&  16.60 (0.12)& 16.23 (0.04)& 6$\star\star$\\   
Apr24  2009      & 2454946.45 &  -- &  -- &  -- &   -- & 16.23 (0.03)&  6 \\  
Apr29  2009      & 2454950.56 &  -- &  -- &  -- &  16.83  (0.09)&  -- &  9$\star$ \\
Apr29  2009      & 2454950.58 &  -- &  -- &  -- &  16.84  (0.09)&  -- &  13$\star$ \\
Apr29  2009      & 2454951.45 &  -- &  -- &  -- &  16.90  (0.09)&  -- &  3$\star$ \\ 
Apr30  2009      & 2454951.74 &  -- &  -- &  -- &  16.93  (0.06)&  -- &  15$\star$ \\ 
May01  2009      & 2454953.37 &  -- &  -- &  -- &  17.04  (0.08)&  -- &  10$\star$\\
May01  2009      & 2454953.42 &  -- &  -- &  -- &  17.03  (0.08)&  -- &  6$\star$\\ 
May02  2009      & 2454954.34 &  -- &  -- &  -- &  17.13  (0.09)&  -- &  3$\star$\\
May02  2009      & 2454954.37 &  -- &  -- &  -- &  17.14  (0.13)&  -- &  10$\star$\\
May02  2009      & 2454954.48 &  -- &  -- &  -- &  17.15  (0.09)&  -- &  17$\star$ \\
May03  2009      & 2454954.78 &  -- &  -- &  -- &  17.19  (0.14)&  -- &  15$\star$ \\
May03  2009      & 2454955.34 &  -- &  -- &  -- &  17.18  (0.06)&  -- &  6$\star$\\
May03  2009      & 2454955.45 &  -- &19.41 (0.13)& 17.94 (0.06)&  17.16 (0.03)& 16.66 (0.02)&  20 \\  
May04  2009      & 2454955.53 &20.85 (0.16) &19.46 (0.03)& 17.97 (0.01)&  17.17 (0.01)& 16.72 (0.01)&  18 \\
May04  2009      & 2454956.39 &  -- &  -- &  -- &  17.29  (0.10)&  -- &  3$\star$ \\
May04  2009      & 2454956.39 &  -- &  -- &  -- &  17.30  (0.07)&  -- &  10$\star$ \\
May05  2009      & 2454957.38 &  -- &  -- &  -- &  17.43  (0.08)&  -- &  10$\star$ \\
May05  2009      & 2454957.42 &  -- &  -- & 18.04 (0.13)&   -- & 16.74 (0.06)& 6 \\ 
May06  2009      & 2454957.60 &  -- &  -- &  -- &  17.44  (0.12)&  -- &  9$\star$ \\
May06  2009      & 2454958.36 &  -- &  -- &  -- &  17.46  (0.07)&  -- &  10$\star$ \\
May06  2009      & 2454958.40 &  -- &  -- &  -- &  17.46  (0.09)&  -- &  3$\star$ \\
May07  2009      & 2454959.35 &  -- &  -- &  -- &  17.47  (0.09)&  -- &  6$\star$\\
May08  2009      & 2454959.54 &  -- &  -- &  -- &  17.51  (0.09)&  -- &  17$\star$ \\
May08  2009      & 2454959.64 &  -- &  -- &  -- &  17.51  (0.20)&  -- &  15$\star$ \\
May10  2009      & 2454961.56 &  -- &  -- &  -- &  17.56  (0.18)&  -- &  6$\star$\\
May10  2009      & 2454962.37 &  -- &  -- &  -- &  17.56  (0.09)&  -- &  17$\star$ \\ 
May10  2009      & 2454962.39 &  -- &  -- &  -- &  17.55  (0.05)&  -- &  3$\star$ \\
May13  2009      & 2454964.66 & $>$20.23 &19.83 (0.13)& 18.42 (0.06)&  17.55  (0.03)& 16.99 (0.03)&  21 \\
May15  2009      & 2454967.36 &  -- &  -- &  -- &  17.61  (0.07)&  -- &  10$\star$ \\
May16  2009      & 2454968.36 &  -- &  -- &  -- &  17.64  (0.11)&  -- &  3$\star$ \\
May17  2009      & 2454968.57 &  -- &  -- &  -- &  17.64  (0.11)&  -- &  7$\star$ \\  
May18  2009      & 2454970.35 &  -- &  -- &  -- &  17.66  (0.10)&  -- &  6$\star$ \\ 
May18  2009      & 2454970.36 &  -- &  -- &  -- &  17.67  (0.10)&  -- &  10$\star$ \\
May20  2009      & 2454971.52 & $>$21.08 &20.04 (0.09)& 18.58 (0.03)&  17.69 (0.03)& 17.11 (0.02)&  21 \\
May20  2009      & 2454972.35 &  -- &  -- &  -- &  17.68  (0.06)&  -- &  6$\star$\\  
May25  2009      & 2454977.34 &  -- &  -- &  -- &  17.70  (0.09)&  -- &  17$\star$ \\ 
May27  2009      & 2454979.37 &  -- &  -- & 18.63 (0.22)&  17.73  (0.10)& 17.20 (0.14)&  16 \\
May31  2009      & 2454982.59 &  -- &20.06 (0.29)& 18.69 (0.12)&  17.72 (0.07)& 17.21 (0.07)&  20 \\ 
May31  2009      & 2454982.65 &  -- &  -- &  -- &  17.78  (0.24)&  -- &  13$\star$ \\
Jun02  2009      & 2454984.66 &  -- &  -- &  -- &  17.80  (0.21)&  -- &  15$\star$ \\
Jun06  2009      & 2454988.50 &  -- &  -- &  -- &  17.83  (0.22)&  -- &  9$\star$ \\
Jun09  2009      & 2454991.65 &  -- &  -- &  -- &  17.86  (0.09)&  -- &  13$\star$ \\
Jun11  2009      & 2454993.58 &  -- &  -- &  -- &  17.86  (0.18)&  -- &  9$\star$ \\
Jun12  2009      & 2454994.62 &  -- &  -- &  -- &  17.86  (0.21)&  -- &  9$\star$ \\
Jun12  2009      & 2454994.72 &  -- &  -- &  -- &  17.89  (0.24)&  -- &  15$\star$ \\
Jun26 2009      & 2455009.38 &  -- &  -- & 19.05 (0.11)&  18.06  (0.04)& 17.45 (0.03)&  20 \\
Jul01  2009      & 2455013.67 &  -- &  -- &  -- &  18.09  (0.31)&  -- &  15$\star$ \\ \hline
\\
\caption{continued.}\\ \hline\hline
Date & $JD$ & $U$ & $B$ & $V$ & $R$ & $I$ & Instrument \\ \hline
Jul18  2009      & 2455031.39 &  -- &  -- & 19.35 (0.10)&  18.26  (0.04)& 17.68 (0.04)&  20 \\ 
Jul29  2009      & 2455042.36 &  -- &  -- & 19.43 (0.11)&  18.35  (0.07)& 17.74 (0.05)&  16 \\ 
Aug12  2009      & 2455056.35 &  -- &20.54 (0.20)& 19.54 (0.15)&  18.52  (0.06)& 17.92 (0.05)&  20 \\
Aug19  2009      & 2455063.37 &  -- &  -- & 19.57 (0.12)&   --  &  -- &  16 \\
Aug20  2009      & 2455064.32 &  -- &  -- &  -- &  18.56  (0.08)&  -- &  16 \\ 
Aug28  2009      & 2455072.34 &  -- &20.65 (0.32)& 19.71 (0.16)&  18.67  (0.11)& 18.12 (0.08)& 20 \\ 
Sep04  2009      & 2455079.34 &  -- &20.70 (0.21)& 19.76 (0.15)&  18.72  (0.14)& 18.16 (0.13)& 20 \\  
Nov19  2009      & 2455154.69 &  -- & --  & 20.29 (0.27) & 19.35 (0.10) & 18.94 (0.04) & 20\\
Jan13  2010      & 2455209.60 &  -- & --  & 20.87 (0.12) & 19.93 (0.09) & 19.57 (0.08) & 18\\
Jan24  2010      & 2455220.69 &  -- & --  &  -- &  20.06 (0.16) & --  & 18$\star$\\
Feb24  2010      & 2455251.50 &  -- & --  & 21.23 (0.37) & 20.30 (0.25) & 20.04 (0.23) & 18 \\  
Mar29  2010      & 2455285.43 &  -- &22.05 (0.45)& 21.55 (0.18) & 20.64 (0.28) & 20.50 (0.12) & 21 \\  
\hline
\end{longtable}
\begin{flushleft} 1 = 0.36-m C14 reflector + Apogee AP7 CCD camera (Obs. T. Boles, Coddendham Observatory, Suffolk,  UK);\\
2 = 0.50-m Telescope + Apogee U6 CCD camera (Obs. M. Yoshida, Akeno Observatory/ICRR, Yamanashi, Japan); \\
3 = 0.28-m C11 reflector + SBIG ST-8XME Kaf1602E CCD camera (Obs. J. Nicolas, Vallauris, France); \\
4 = 0.28-m C11 reflector + SBIG ST-8XME Kaf1602E CCD camera (Obs. J. M. Llapasset, Perpignan, France); \\
5 = 0.36-m C14 reflector + Apogee AP7 CCD Camera (Obs. O. Trondal, Groruddalen, Oslo, Norway); \\
6 = 0.36m Meade LX200 Telescope + SBIG ST-9XE CCD camera (Obs. E. Prosperi, Osservatorio Astronomico di Castelmartini, Larciano, Pistoia, Italy); \\
7 = 0.3-m Takahashi Mewlon 300 + SBIG ST-8E NABG camera (obs. W. Wells, Gras-002, New Mexico, USA);\\
8 = 2.0-m Faulkes Telescope North + EM01 (Faulkees Telescope Archive - Las Cumbres Observatory, Mt. Haleakala, Hawaii Islands, USA);\\
9 = 0.35-m Bradford Robotic Telescope + FLI MaxCam CM2-1 camera with E2V CCD47-10 (Obs. G. Hurst, Tenerife Observatory, Canary Islands, Spain);\\
10 = 0.20-m C8 reflector + SBIG ST-9 Kaf0261 CCD camera (obs. A. San Segundo Delgado, Observatorio El Guijo, Galapagar, Madrid, Spain);\\
11 = 0.25-m Newton Telescope + Meade DSI Pro camera with Sony EXView HAD CCD (Obs. R. Mancini and F. Briganti, Associazione Astronomica Isaac Newton, Stazione di Gavena, Cerreto Guidi, Italy);\\
12 = 0.40-m reflector + DTA camera with Kodak Kaf0260 CCD (G. Iacopini, Osservatorio della Tavolaia, Associazione Astronomica Isaac Newton, Santa Maria a Monte, Pisa, Italy);\\
13 = 0.25-m Meade 10" LX200 Telescope + SBIG ST-8XME Kaf1602E CCD camera (Obs. A. D. Cason, Dawsonville, Georgia, USA);\\ 
14 = 0.5-m Newton-Cassegrain Telescope + Hi-Sis 44 CCD camera (Obs. A. Dimai, Osservatorio di Col Driusci\`e, Cortina, Italy);\\
15 = 0.3-m Takahashi Mewlon 300 + FLI IMG1024 DM camera (obs. S. Howerton, Gras-001, New Mexico, USA);\\ 
16 = 1.82-m Copernico Telescope + AFOSC (INAF - Osservatorio Astronomico di Asiago, Mt. Ekar, Asiago, Italy); \\
17 = 0.30-m Meade LX200 Telescope + SBIG ST-10XME CCD camera (Obs. E. Prosperi, Skylive Remote Facility, Osservatorio B40 Skylive, Pedata, Catania, Italy);\\
18 = 2.56-m Nordic Optical Telescope + ALFOSC (La Palma, Canary Islands, Spain);\\
19 = 0.7-m Ritchey-Chr\'etien Telescope + Apogee Alta U9000 camera with Kodak Kaf-09000 CCD (Obs. A. Englaro, I. Bano and G. Cetrulo, Osservatorio Astronomico di Polse di Cougnes, Zuglio, Udine, Italy);\\
20 = 2.2-m Calar Alto Telescope + CAFOS (German-Spanish Astronomical Center, Andaluc\'ia, Spain); \\
21 = 3.58-m Telescopio Nazionale Galileo + Dolores (Fundaci\'on Galileo Galilei - INAF, La Palma, Canary Islands, Spain).\\
\end{flushleft}
}

The $R$-band absolute light curve of SN 2009E is shown in Figure \ref{Fig4}, together with those of a
number of other type II SNe with a broad, delayed light curve peak, similar to that observed for SN
1987A. In analogy with the early discovery of SN 1987A \citep{arn89}, 
also SN 2009E was likely caught soon after the core-collapse. 
Marginal evidence of the initial sharp peak due to the shock breakout is possibly visible (Figure \ref{Fig4}, panel C), 
although the large error bars of the unfiltered amateur images make this finding rather uncertain. 

Nevertheless, we can constrain the explosion epoch of SN 2009E with a small uncertainty through a comparison
with the early photometric evolution of SN 1987A. Hereafter we will adopt $JD$=2454832.5$^{+2}_{-5}$ 
as the best estimate for the core-collapse epoch of SN 2009E. 
The distances and reddenings used for our sample of 1987A-like events shown in Figure \ref{Fig4} 
are listed in Tables \ref{Tab4} and \ref{Tab5}. SN 2009E
is the faintest SN in this sample at all the epochs. In particular, its peak magnitude $M_R \sim -$16.2
is about 0.4 mag fainter than that of SN 1987A and its $^{56}$Ni mass, as derived from the
luminosity of the radioactive tail relative to that of SN 1987A, is 0.040$^{+0.015}_{-0.011}$ M$_\odot$, the lowest among the objects in our sample.

Another remarkable property of SN 2009E is that its $R$-band light curve peaks at about +96d after
core-collapse, significantly later than SN 1987A ($\sim$ +84 days). 
This suggests different ejecta parameters for SN 2009E (e.g. smaller initial radius, lower expansion velocity or a more massive envelope, see Section \ref{model}). Interestingly, fainter SNe in our sample seem to reach
peak magnitude later than the more luminous ones (see also Appendix \ref{appB}).

\subsection{Spectra} \label{sp}
\begin{table*}
\caption{Spectroscopic observations of SN~2009E.}\label{Tab3}
\centering
\begin{tabular}{ccccccc}
\hline \hline
Date & Average $JD$ & Phase$^\star$ & Instrumental configuration & Exposure (s) & Resolution (\AA)$^\medbullet$ & Range (\AA) \\ \hline
Mar24, 2009 & 2454915.40 & +82.9 & Ekar1.82m+AFOSC+gr.4 & 3600 & 24 & 3490--7790 \\
Mar25, 2009$^\ddag$ & 2454915.54 & +83.0 & Asiago1.22m+B$\&$C+g600 & 3600 & 6.1 & 5090--7500 \\
Apr01, 2009 & 2454922.52 & +90.0 & NOT+ALFOSC+gr4 & 1200 & 13 & 3360-9100 \\ 
Apr12, 2009 & 2454934.48 & +102.0 & NOT+ALFOSC+gr4 & 1200 & 16 & 3230-9060 \\ 
Apr16, 2009 & 2454938.42 & +105.9 & CAHA2.2m+CAFOS+g200 & 1792 & 9.5 & 3700-10670 \\ 
May04, 2009 & 2454955.57 & +123.1 & NOT+ALFOSC+gr4 & 2700 &  13  & 3350-9120 \\
May13, 2009 & 2454964.62 & +132.1 & TNG+LRS+LRR & 3600 & 12 & 5050-9520  \\
May20, 2009 & 2454971.55 & +139.1 & TNG+LRS+LRR & 3600 & 12 & 4960-9560 \\
Jun26, 2009$^\ddag$ & 2455009.42 & +176.9 & CAHA2.2m+CAFOS+g200 & 2$\times$2400  & 9.5 & 4100-9100 \\
Jul18, 2009 & 2455031.44 & +198.9 & CAHA2.2m+CAFOS+g200+GG495 & 2$\times$3600 & 13 & 4800-9690  \\
Aug15, 2009$^\dag$  & 2455058.84 & +226.3 & CAHA2.2m+CAFOS+g200 & 2$\times$3600 & 12 & 3880-10670 \\
Jan22, 2010 & 2455218.64 &  +386.1 & NOT+ALFOSC+gr5 & 3$\times$1200 &  14 & 5000--10000 \\
Jan24, 2010 & 2455220.70 &  +388.2 & NOT+ALFOSC+gr4 & 3$\times$1200 &  12 & 3550--9080 \\
\hline
\end{tabular}

$^\star$ Days from explosion ($JD$=2454832.5);
$^\medbullet$ As estimated from the full width at half maximum of isolated night sky lines. 
$^\ddag$ Poor signal-to-noise spectrum; $^\dag$ Weighted average of 2 spectra obtained on August 14th and 15th.
\end{table*}

   \begin{figure}
   \centering
   \includegraphics[angle=0,width=9.2cm]{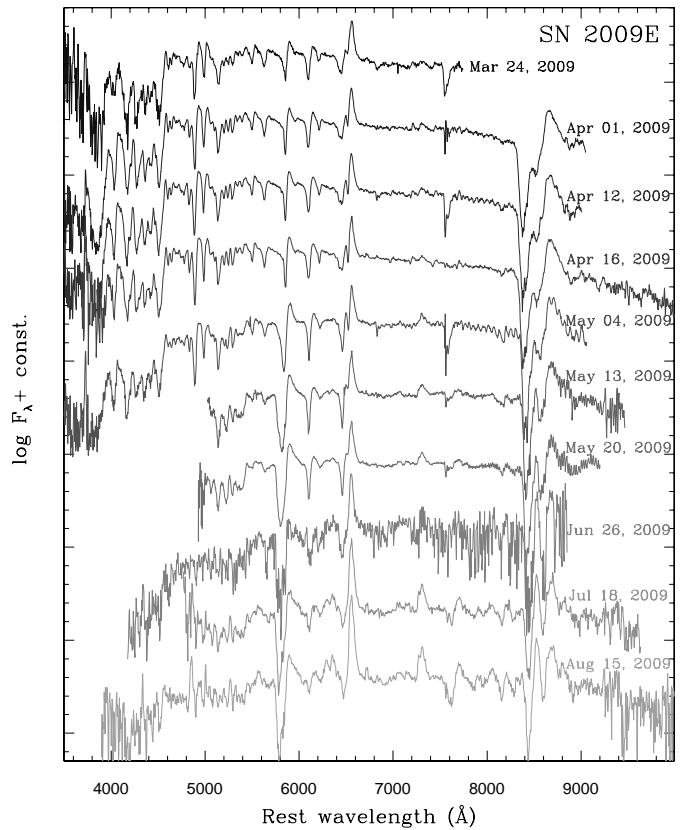}
   \caption{Spectral sequence for SN 2009E, starting from about 80 days after core-collapse. }
              \label{Fig5}
    \end{figure}

As mentioned above, the spectroscopic classification of SN 2009E was performed about 80 days after 
the SN discovery. To our knowledge, no spectra exist to witness the early time evolution of this object.
After the SN classification, we started a regular spectroscopic monitoring covering the SN evolution
from near maximum light to the nebular phase. Spectra have been reduced in the IRAF environment using standard prescriptions.
After the optimal extraction, the SN spectra were wavelength calibrated using arc lamp spectra and then flux calibrated 
using flux standard stars observed in the same nights as the SN. A final check with photometry was performed in order to fine tune the flux calibration of the spectra, with a remaining uncertainty in the flux calibration of less than 10$\%$. 
Basic information on the spectra collected for SN 2009E is listed in Table \ref{Tab3}, while the spectral sequence until August 15, 2009 is shown in Figure \ref{Fig5}. 

\begin{figure}
   \centering
   \includegraphics[angle=0,width=9.2cm]{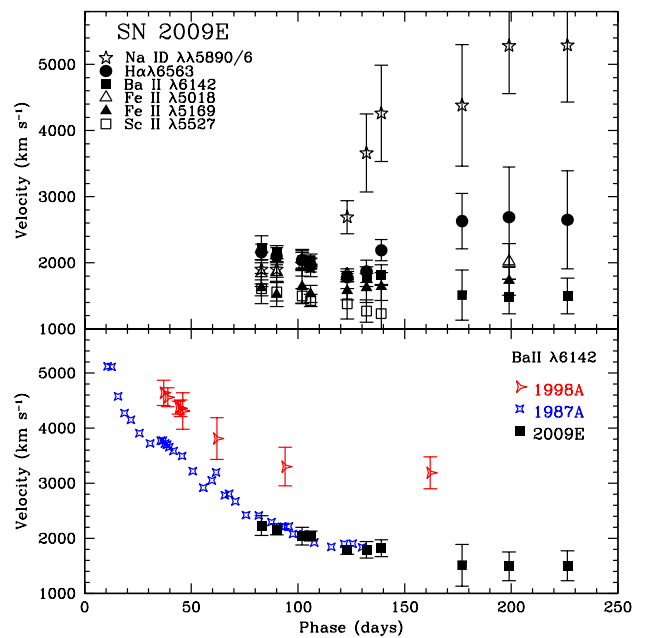}
   \caption{Top: Evolution of the expansion velocities for the principal lines in the spectra of SN 2009E.
   Bottom: comparison of the velocities derived for the Ba II $\lambda$6142 line in SNe 2009E, 1987A and 1998A.}
              \label{Fig6}
    \end{figure}

Since the earliest spectrum 
was obtained about 83 days after the core-collapse, the continuum is quite red, indicative of relatively low temperatures
(T$_{bb} \approx$ 6000 K). All the photospheric spectra show a clear  flux deficit at the blue wavelengths (below 4500~\AA) that is due to line blanketing from increasingly strong metal lines.
These spectra are rich in narrow P-Cygni lines indicating remarkably low velocity of the expanding ejecta. 
From the position of the minima of the 
Sc II $\lambda$6245 and $\lambda$5527 lines, we measured v$_{ph} \approx$ 1600 km s$^{-1}$ in our earliest spectrum
and about 1000 km s$^{-1}$ at the end of the photospheric phase (Figure \ref{Fig6}). In the same period, the velocity of the 
unblended Ba II $\lambda$6142 line declines from 2200 km s$^{-1}$ to 1850 km s$^{-1}$.

\begin{figure}
   \centering
   \includegraphics[angle=0,width=9.2cm]{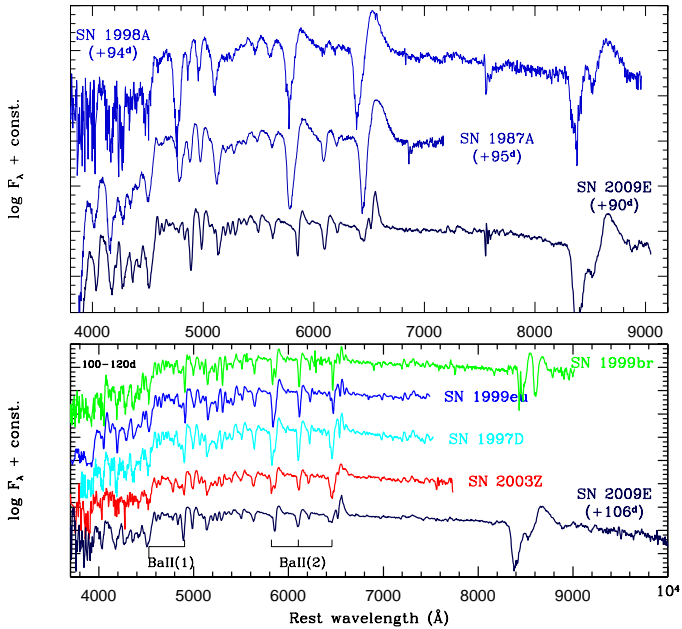}
   \caption{Top: A spectrum of SN 2009E obtained on Apr 1st, 2009 is compared with spectra of 
   SNe 1998A \citep{pasto05} and 1987A \citep{pun95} at a similar phase. 
    Bottom: a post-maximum spectrum of SN 2009E (Apr 16th, 2009) is compared with spectra of underluminous type IIP
   SNe obtained around the end of the plateau phase \citep{tura98,pasto04}.}
              \label{Fig7}
    \end{figure}

In Figure \ref{Fig7}, a late time photospheric spectrum of SN 2009E (phase about 3 months after explosion) is compared with spectra of SN 1987A and the 1987A-like event SN 1998A \citep[][; see also Figure \ref{Fig7}, top panel]{pasto05}, and with spectra of faint SNe IIP \citep[][see Figure \ref{Fig7}, bottom panel]{tura98,ben01,pasto04,pasto09}.  While the strong Ba II lines are a common
feature both for the spectra of SN 1987A and under-luminous, $^{56}$Ni-poor type IIP SNe, the narrowness of the spectral lines in SN 2009E
is more reminiscent of 1997D-like events \citep{tura98,pasto04}.
 
   \begin{figure}
   \centering
   \includegraphics[angle=0,width=9.2cm]{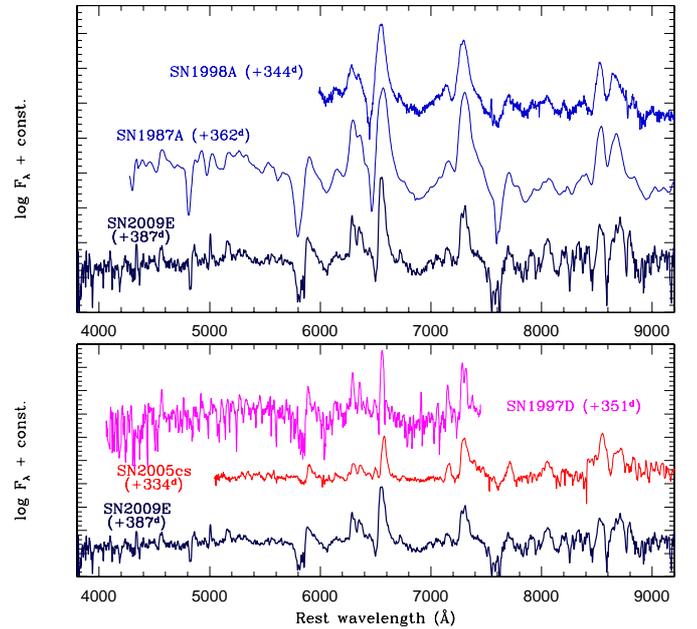}
   \caption{Top: A nebular spectrum of SN 2009E is 
   compared with spectra of SNe 1998A \citep{pasto05} and 1987A \citep{pun95} at a similar phase ($\sim$ 1 year).
   The nebular spectrum of SN 2009E was obtained averaging spectra
   of 2010 Jan 22nd and 24th. 
   Bottom: the nebular spectrum of  SN 2009E is compared with spectra of underluminous type IIP
   SNe obtained about 1 year after core-collapse \citep{ben01,pasto09}.}
              \label{Fig8}
    \end{figure}

Together with quite a weak H$\alpha$, we clearly identify prominent lines of
Na ID, Ca II, O I, Fe II, Ti II, Sc II and Cr II [see \cite{pasto06} for a detailed line identification for underluminous SNe IIP during the photospheric phase, and \cite{ben01} for the nebular phase].
The Ba II lines (multiplet 2) are among the strongest spectral features. This is a common characteristic of both SN 1987A \citep{wil87,maz92,maz95} and faint SNe IIP \citep{pasto04}. The presence of prominent Ba II lines
in the photospheric spectra of SN 1987A was  interpreted by \citet{tsu01} as a signature of 
an over-abundance of Ba in the outer layers of the SN \footnote{Note that \citet{utr05} favoured 
 time-dependent ionization effects to explain the presence of prominent Ba II lines in photospheric spectra of SN 1987A.}. However, the presence of prominent Ba II features in 
underluminous SNe IIP can also be explained invoking temperature effects.
The evolution of Na ID is puzzling. The absorption component becomes broader with time
(see Figure \ref{Fig5} for the evolution of this spectral feature from 2009 April 16 to May 13).
A similar behaviour, although less extreme, is observed also for H$\alpha$.
This effect is probably due to line blending, with the Ba II $\lambda$5854 and $\lambda$6497 lines 
(and possibly of other metal features) having an increasing strength relative to Na ID and H$\alpha$ respectively.
The evolution of the expansion velocities, as deduced from the minima of the P-Cygni of H$\alpha$, Na I, Ba II,
Fe II and Sc II, is shown in Figure \ref{Fig6}. Metal lines show a monotonic decline and velocities at the end of the plateau
phase (1200$-$1800 km s$^{-1}$) that are only marginally higher than those observed in sub-luminous type IIP SNe 
\citep[800$-$1200 km s$^{-1}$, see e.g.][]{tura98,pasto04,pasto09}, whilst H$\alpha$ and Na ID display an opposite trend.
Since May (around 4 months past explosion, Figure \ref{Fig5}), the classical nebular doublets of [Ca II] $\lambda\lambda$ 7291-7323 
and [O I] $\lambda\lambda$ 6300-6364 become visible, and increase
in strength with time.  

A late time spectrum of SN 2009E was obtained at the Nordic Optical Telescope (NOT) about 1 year after explosion (see Figure \ref{Fig8}). 
Although the transition toward a purely nebular appearence is not complete (e.g. broad P-Cygni features of Na ID, O I $\lambda$7774 and 
Ca II NIR are still visible), the spectrum is clearly dominated by prominent H$\alpha$ in emission, [O I] $\lambda\lambda$6300,6364
and [Ca II] $\lambda\lambda$7291,7323. In addition, a number of [Fe II] features are detected. In this phase the spectrum of SN 2009E
more closely resembles those of  SN 1987A and SN 1998A (Figure \ref{Fig8}, top), although the spectral lines are significantly narrower 
than those of other 1987A-like events and are closer to those of sub-luminous, $^{56}$Ni-poor SNe (Figure \ref{Fig8}, bottom).

\section{Constraining the progenitor of SN 2009E} \label{const_pro}

\subsection{The ejected oxygen mass} \label{OI}

The strength of the [O I] $\lambda\lambda$6300,6364 lines in the nebular spectra of core-collapse SNe can be used to roughly estimate the O mass ejected 
in the SN explosion \citep{uom86,li92,chu94}. This method has been applied for determining the O masses
for a few type IIP SNe \citep[see][ and references therein]{mag10}. We first make an attempt to estimate the minimum O
mass needed to produce the [O I] $\lambda\lambda$ 6300-6364 feature with the relation 
between the O mass and the total flux of the doublet presented by \citet{uom86}.
Using the nebular spectrum of SN 2009E, we measure an observed flux F$_{[OI]} \approx$ 3.35 $\times$ 10$^{-15}$ erg s$^{-1}$ cm$^{-2}$. Accounting for the reddening and distance estimates as mentioned in  Section \ref{hg}, we obtain for SN 2009E a 
minimum O mass in the range between 0.05 M$_\odot$ and 0.22 M$_\odot$ \citep[assuming the extreme values of 3500 K and 4500 K, respectively as the temperature of the O-rich material, like in][]{mag10}. 
As a comparison, \cite{mag10} derived a minimum O mass range for SN 1987A of 0.3$-$1.1 M$_\odot$. The only other 1987A-like object with nebular spectra available is SN 1998A \citep{pasto05}. Using the spectrum at phase $\sim$ 344 days (see Figure \ref{Fig8}, top), we obtain an O mass in the range 0.24$-$1.01 M$_\odot$.

In order to improve the O mass estimate, we adopt the method already used in \cite{elm03}, which is based on the fact that the luminosity of the [O I] doublet in the nebular phase is known to be powered by the $\gamma$-rays produced in the radioactive 
decays of $^{56}$Co to $^{56}$Fe, and deposited in the O-rich ejecta.
Since the mass of O in SN 1987A is reasonably well-known \citep[1.2$-$1.5 M$_\odot$,][]{li92,chu94,koz98},
we can derive the O masses of SNe 2009E and 1998A via comparison with the O mass of SN 1987A.
We use the relation linking the O mass with the luminosities of the [O I] doublet\footnote{We note that the luminosity of the [O I] doublet 
in SN 1987A and type IIP SNe remains almost constant over a long period during the nebular phase
\protect\citep[between $\sim$ 200-400 days, see e.g. Figure 8 in][]{elm03}.} and of the radioactively powered
light curve tail, computed at similar phases \citep[][]{elm03}:

\begin{equation} \label{eq1}
L_{[OI]} = \eta_O \frac{M_O}{M_{exc}} L_{^{56}Co},
\end{equation}

\noindent where L$_{[OI]}$ and L$_{^{56}Co}$ are the luminosities of the [O I] doublet and of the radioactive tail (respectively),
M$_O$ is the mass of O, M$_{exc}$ represents the whole excited mass in which the radioactive energy is deposited,
and $\eta_O$ is the efficiency of the transformation of the deposited energy into [O I] line radiation.
Assuming that M$_{exc}$ and $\eta_0$ are similar in all 1987A-like events and noting that (since the late light curve of SN 2009E follows the $^{56}$Co decay rate) 

\begin{equation}
\frac{L_{^{56}Co}(1987A)}{L_{^{56}Co}(SN)} \propto \frac{M_{^{56}Ni}(1987A)}{M_{^{56}Ni}(SN)},
\end{equation}
from Equation \ref{eq1}, we obtain:
\begin{equation} \label{eq3} 
M_O(SN) = M_O(1987A) \times   \frac{L_{[OI]}(SN) \times M_{^{56}Ni}(1987A)}{L_{[OI]}(1987A) \times M_{^{56}Ni}(SN)}.   
\end{equation}

Now we have all the ingredients needed to estimate the O mass in SNe 2009E and 1998A from Equation \ref{eq3} using the information 
available for SN 1987A \citep{li92,chu94}.
For SN 2009E we obtain M$_O \approx$ 0.60$-$0.75 M$_\odot$ (using as O mass for SN 1987A the extreme values of M$_{O,min}$ = 1.2 M$_\odot$ and 
M$_{O,max}$ = 1.5 M$_\odot$, respectively),
while for SN 1998A we obtain M$_O \approx$  1.18$-$1.48 M$_\odot$. 
Therefore, with this method, we determine for the under-luminous SN 2009E an ejected mass 
of O that is a factor of 2 smaller than that of SN 1987A (note that for SN 2009E also the $^{56}$Ni mass is a factor of $\sim$2 smaller). 
In the case of SN 1998A we obtain an O mass that matches the values of SN 1987A, despite the former object ejected 
more $^{56}$Ni.

According to the models presented by \cite{woo95} and \cite{thi96}, a star with main sequence mass of about 18$-$20 M$_\odot$ 
produces 1.2$-$1.5 M$_\odot$ of oxygen and a $^{56}$Ni mass of 0.07-0.1 M$_\odot$. This is in excellent agreement with the values observed in SNe 1987A and 1998A.
It is more difficult to provide an interpretation for the abundances displayed by SN 2009E 
(both M$_O$ and M$_{^{56}Ni}$ are smaller by a factor 2 than those of SN 1987A), although we may tentatively guess that the 
lower O and $^{56}$Ni masses would point toward a lower-mass main sequence star. 
We remark that the above estimates have been obtained under the simplifying assumption that M$_{exc}$ and $\eta_O$ were the same for 
SNe 1987A, 1998A and 2009E. This is not necessarily true, since we can reasonably expect a range of values for these parameters in 1987A-like events,
and this would significantly affect our O mass estimates.

\subsection{Modelling the data of SN 2009E} \label{model}

Other physical properties of SN 2009E (namely, the ejected 
mass, the progenitor initial radius and the explosion energy) have been estimated 
by performing a model vs. data 
comparison using two  codes that compute the bolometric 
light curve, the evolution of the ejecta velocity and the continuum temperature 
at the photosphere of a SN. The first code, based on a simplified semi-analytic treatment
of the ejecta evolution \citep{zam03}, is used to perform a preparatory 
study in order to constrain the parameter space. The second code, that includes 
a more accurate treatment of radiative transfer and radiation hydrodynamics 
\citep{pumo10,pumo11}, computes the tighter grid 
of models needed for the final comparison. This is similar to the procedure 
adopted for SN 2007od in \cite{ins11}.

   \begin{figure*}
   \centering
   \includegraphics[angle=270,width=18cm]{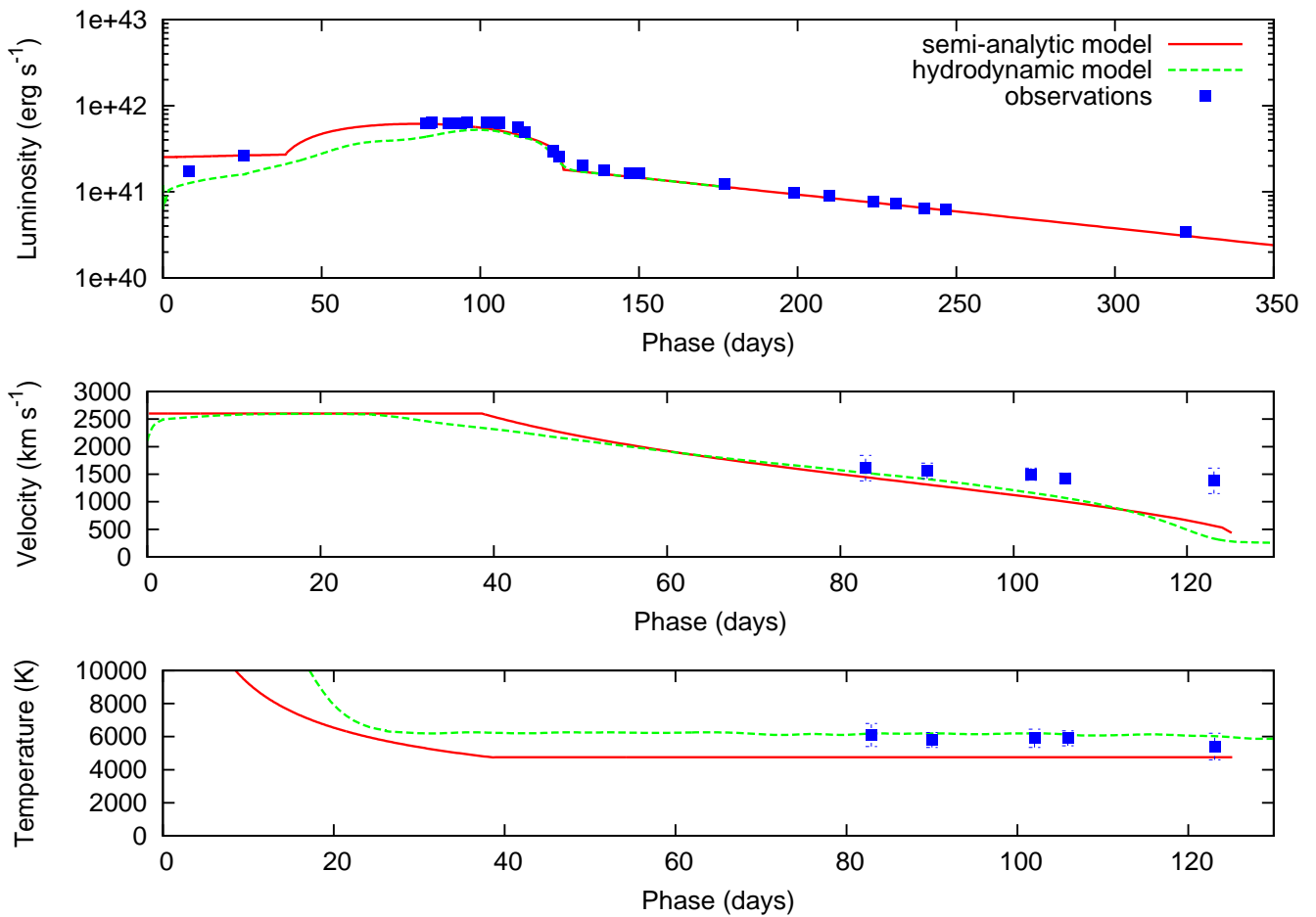}
   \caption{Comparison of the evolution of the main observables of SN 2009E with the best-fit model 
computed with the relativistic radiation-hydrodynamics code (total energy 0.6 foe, 
initial radius $7 \times 10^{12}$ cm, envelope mass  19 M$_\odot$). The best-fit model 
computed with the semi-analytic code (total energy 1.3 foe, initial radius 
$6 \times 10^{12}$ cm, envelope mass $26 M_\odot$) is also shown.
Top, middle, and bottom panels show respectively the bolometric light curve, the photospheric velocity, 
and the photospheric temperature as a function of time. As a tracer for the photospheric 
velocity we measured the positions of the minimum of the P-Cygni profiles of the Sc II lines.}
              \label{Fig9}
    \end{figure*}

The bolometric light curve was computed from the observed
multicolor light curves using the assumptions on extinction and distance
modulus reported in Section~\ref{hg}. We have adopted the explosion epoch estimated in Section~\ref{lc} ($JD$ $= 2454832.5^{+2}_{-5}$). 
Finally, we have assumed that SN 2009E had the same color evolution as SN 1987A (this is a reasonable assumption, see e.g. Figure \ref{Fig3}), and hence

\begin{equation}
L(2009E) = L(1987A) \times \frac{L_{BVRI}(2009E)}{L_{BVRI}(1987A)},
\end{equation}

\noindent where $L$ and $L_{BVRI}$ are the bolometric and quasi-bolometric 
($BVRI$) luminosities of the two SNe \citep[bolometric data for SN 1987A were taken from][]{cat87,cat88}. 

The best $\chi^{2}$ fit to the bolometric light curve of SN 2009E obtained with the
full radiation-hydrodynamics code returned values of a total (kinetic plus thermal) energy (E)
of 0.6 foe, an initial radius (R$_0$) of $7 \times 10^{12}$ cm, and an envelope mass (M$_{ej}$) of $19 M_\odot$ (see Figure 
\ref{Fig9}). As a comparison, the fit with the semi-analytic code 
gave E = 1.3 foe, R$_0 = 6 \times 10^{12}$ cm, and 
M$_{ej} = 26$ M$_\odot$ (see again Figure \ref{Fig9}).  
The parameters inferred from the two best-fits are in reasonable agreement. 
The envelope mass (and consequently the total energy) are slightly overestimated by the semi-analytic code as a 
consequence of the different mass distribution. Indeed, given the assumption
of uniform density throughout the ejecta, the semi-analytic code tends to gather
more mass (and hence more energy) in the external, faster-moving layers. 
The $^{56}$Ni distribution also affects the mass estimate because, as a consequence of the 
simplified uniform distribution adopted in the semi-analytic code, less energy is released at 
the end of the photospheric phase and, hence, a more massive envelope is needed to sustain the 
recombination phase.

The ejected $^{56}$Ni mass of 0.039 M$_\odot$ that is adopted in the semi-analytic code is in an excellent agreement with that estimated in Section~\ref{lc}, whilst the one inferred from the radiation-hydrodynamic code is slightly higher (0.043 M$_\odot$).
Since the latter code accounts also for the amount of material which (eventually) falls back onto the central remnant during the post-explosive evolution, the numerical simulations need a larger initial $^{56}$Ni mass to reproduce the observed late-time light curve of SN 2009E.
 
In Figure \ref{Fig9} the evolutions of the photospheric velocity and temperature are also shown.
The agreement between models and observations in the late photospheric phase is good, with possibly some discrepancy in matching the photospheric velocities at later times 
($\gtrsim$ 4 months). 
The decline of the model velocity profile is faster than that inferred from the observations, probably because 
some extra radioactive decay heating occurs in the inner part of the ejecta of SN 2009E as a consequence 
of a different (i.e. more centrally condensed) distribution of $^{56}$Ni.

\begin{table*}
\caption{Parameters of the explosion sites of the 1987A-like SNe. In column 3 we report
the morphologic types of the host galaxies, in columns 4 and 5 the adopted distance moduli
and Galactic extinctions from \protect\cite{sch98}, in column 6 the 
$B$-band absolute magnitudes of the galaxies, 
in column 7 the ratios between the deprojected position of the SN  and r$_{25}$ computed following \cite{hak09}, 
in column 8 the oxygen abundances from Pilyugin's relation \protect\citep[computed at 0.4r$_{25}$,][]{p04},
in column 9 the corrected oxygen abundances from Pilyugin's relation computed
at the deprojected SN distance.}\label{Tab4}
\centering
\begin{tabular}{ccccccccc}
\hline \hline
SN & Host galaxy & Type$^{1}$ &$\mu$$^{2}$  & A$_{B,MW}$ & $M_B$  & d$_{SN}^0$/r$_{25}$ & P04 & P04 corr. \\ \hline
1909A & NGC 5457 & SABc & 29.34$^{3}$ & 0.037 & -21.01 & 1.11$^{9}$ & 8.59 & 8.30$^{10}$ \\
1982F & NGC 4490 & SBcd & 29.92 & 0.093 & -20.25 & 0.23 & 8.53 & 8.60$^{11}$ \\ 
1987A & LMC      & SBm & 18.50$^{4}$ & 0.324 & -18.02 & - & 8.31 & - \\ 
1998A & IC 2627  & SABc & 32.19 & 0.518 & -20.03 & 0.66 & 8.51 & 8.41 \\ 
1998bt& A132541-2646 & -  & 36.41$^{5}$ & 0.244 & -13.25 & - & 7.98 & - \\  
2000cb& IC 1158  & SABc & 32.68$^{6}$ & 0.491 & -19.34 & 0.93 & 8.46 & 8.24 \\ 
NOOS-005 & 2MASX J05553978-6855381 & S & 35.46 & 0.656$^{8}$ & -20.01 & 1.26 & 8.51 & 8.17 \\ 
2004em& IC 1303  & Sc  & 34.07 & 0.466 & -19.73 & 0.96 & 8.49 & 8.26 \\ 
2005ci & NGC 5682  & Sb & 32.73 & 0.141 & -18.07 & 0.66 & 8.36 & 8.26 \\
2006V & UGC 6510 & SABc  & 34.36$^{7}$ & 0.125 & -20.97 & 1.31 & 8.59 & 8.22 \\ 
2006au & UGC 11057 & Sc & 33.38$^{7}$ & 0.742 & -20.64 & 0.90 & 8.56 & 8.36 \\ 
2009E & NGC 4141 & SBc & 32.38 & 0.086 & -17.75 & 0.68 & 8.33 & 8.22 \\   
\hline
\end{tabular}
\begin{flushleft}
$^{1}$ Morphologic type as quoted by HyperLeda.\\
$^{2}$ Unless otherwise specified, distance moduli are computed from HyperLeda's v$_{Vir}$, adopting H$_0$ = 72 km s$^{-1}$ Mpc$^{-1}$.\\
$^{3}$ Distance computed from the tip of the red giant branch \protect\citep[TRGB,][]{riz07}. Cepheid distances are found to be 
too sensitive to metallicity. The metallicity correction is even more crucial in the case of NGC 5457 where  
inner metal-poor and outer metal-rich cepheids give uncorrected distance moduli that may differ by 0.3-0.4 mags \protect\citep{sah06}.\\
$^{4}$ For LMC we adopted a cepheid distance, well consistent with that derived from the TRGB \protect\citep{sak04,riz07}.\\
$^{5}$ SN 1998bt was discovered in the course of the Mount Stromlo Abell cluster supernova search \protect\citep{ger04}. Unfortunately, no spectroscopic classification exists for this SN.\\
According to \protect\cite{ger04}, there is a very faint host galaxy at the position of SN 1998bt, but no spectrum was ever obtained
of this galaxy. We assume that the galaxy hosting SN 1998bt belongs to a cluster monitored by the search whose 
average redshift is z = 0.046.\\
$^{6}$ Average EPM distance from Table 4.1 of \protect\cite{ham01}.\\
$^{7}$ Distance as in \protect\citet{tad11}, but obtained adopting an Hubble Constant H$_0$ = 72 km s$^{-1}$ Mpc$^{-1}$.\\
$^{8}$ This galaxy is behind LMC, so the extinction reported here
includes both Galaxy and LMC contributions (see text for references).\\
$^{9}$ Computed using the position angle provided by \protect\cite{jar03}. \\
$^{10}$ Other sources give 12+log(O/H) $\approx$ 7.7-7.8 at the position of SN 1909A \protect\citep{ken03,p04,bre07}.\\
$^{11}$ An alternative estimate from \protect\cite{p07} gives 12+log(O/H) = 8.35.\\
\end{flushleft}
\end{table*}

\begin{table*}
\caption{Adopted parameters for 1987A-like SNe. The total extinction is reported in column 2, the $JD$ of explosion
in column 3, the bands in column 4, the $JD$ of the maximum light in column 5, the phase of the broad  maximum (in days past explosion) 
in column 6, the apparent and absolute peak magnitudes in columns 7 and 8 (respectively) and the $^{56}$Ni mass
in column 9. The extinction law of \protect\cite{car89} was used in these estimates. The explosion epochs have been
determined through a comparison with the light curves of SN 1987A. }\label{Tab5}
\centering
\begin{tabular}{ccccccccccc}
\hline \hline
SN &  E$_{tot}$(B-V)$^{a}$ & $JD$$_{expl.}$ & Band & $JD$$_{peak}$ & t-t$_0$ (days) & $m_{peak}$(filter) & $M_{peak}$ & M($^{56}$Ni) & Sources\\ \hline
1909A & 0.009 & 2418310$\pm$30 & $B$ & 2418383$\pm$7 & 73$\pm$31 & 14.47$\pm$0.30 & -15.91 & 0.13M$_\odot$$^{b}$ & 1 \\
1982F & 0.022 & 2445014$\pm$15 & $B$ & 2445087$\pm$4 & 73$\pm$16 & 16.40$\pm$0.02 & -13.61 & -- & 2,3,4\\
      &       &            & $V$ & 2445097$\pm$3     & 83$\pm$15 & 14.95$\pm$0.10 & -15.04 &  & \\
1987A & 0.19  & 2446849.816$^{c}$ & $B$ & 2446932.0$\pm$1.1 & 82.2$\pm$1.1 &  4.52$\pm$0.01 & -14.77 & 0.075M$_\odot$ & 5 \\ 
      &       &            & $V$ & 2446932.2$\pm$1.0 & 82.4$\pm$1.0 &  2.96$\pm$0.01 & -16.13 &       &  \\ 
      &       &            & $R$ & 2446933.6$\pm$1.0 & 83.8$\pm$1.0 &  2.27$\pm$0.01 & -16.67 &       &  \\ 
      &       &            & $I$ & 2446935.2$\pm$0.9 & 85.4$\pm$0.9 &  1.90$\pm$0.01 & -16.88 &       &  \\ 
1998A & 0.120 & 2450801$\pm$4 & $B$ & 2450876.3$\pm$7.4 & 75.3$\pm$8.4 & 17.59$\pm$0.13 & -15.12 & 0.09M$_\odot$ & 6 \\
      &       &               & $V$ & 2450885.4$\pm$8.3 & 84.4$\pm$9.2 & 16.31$\pm$0.09 & -16.28 &      &     \\
      &       &               & $R$ & 2450885.6$\pm$3.9 & 84.6$\pm$5.6 & 15.73$\pm$0.06 & -16.78 &      &     \\
      &       &               & $I$ & 2450889.2$\pm$5.5 & 88.2$\pm$6.8 & 15.37$\pm$0.07 & -17.05 &      &     \\
1998bt & 0.057 & 2450831$\pm$30    & $V$  & 2450905$\pm$6 & 74$\pm$31 & 19.81$\pm$0.05 & -16.79 &  & 7,0 \\
      &       &               & $R$ & 2450908$\pm$8 & 77$\pm$31 & 19.38$\pm$0.08 & -17.18 &  & 8,0 \\
2000cb& 0.112 & 2451653.8$\pm$1.4  & $B^{d}$ &                   &              & 18.23$\pm$0.02 & -14.94 & 0.11M$_\odot$ & 8,9,10 \\
      &       &            & $V$ & 2451714.6$\pm$4.7 & 60.8$\pm$4.9 & 16.58$\pm$0.02 & -16.47 &  &  \\
      &       &            & $R$ & 2451725.2$\pm$2.1 & 71.4$\pm$2.5 & 15.95$\pm$0.02 & -17.03 &  &  \\
      &       &            & $I$ & 2451729.0$\pm$1.4 & 75.2$\pm$2.0 & 15.56$\pm$0.02 & -17.34 &  &  \\
NOOS-005&0.160 & 2452855$\pm$15 & $I$ & 2452939.5$\pm$2.7 & 84.5$\pm$15.2 & 18.22$\pm$0.02 & -17.51 & 0.13M$_\odot$ & 0 \\
2005ci & 0.033? & 2453521$\pm$3 & $unf.$ &  &  & $<$17.5 & $<$-15.3 & & 10 \\ 
2006V$^{e}$  & 0.029 & 2453748$\pm$4 & $B$ & 2453823.7 & 75.7 & 18.29$\pm$0.01 & -16.20 & 0.13M$_\odot$ & 11 \\ 
2006au$^{e}$ & 0.312 & 2453794$\pm$9 & $B$ & 2453865.5 & 71.5 & 18.63$\pm$0.02 & -16.06 & $<$0.07M$_\odot$ & 11 \\  
2009E &  0.04  & 2454832.5$\pm$2.0  & $B$ & 2454919.0$\pm$ 6.5 & 86.5$\pm$6.8 & 18.01$\pm$0.03 & -14.54 & 0.04M$_\odot$ & 0 \\
 &    &   & $V$ & 2454924.9$\pm$ 5.9 & 92.4$\pm$6.2 & 16.76$\pm$0.04 & -15.74 & &  \\
 &    &   & $R$ & 2454928.3$\pm$ 2.8 & 95.8$\pm$3.4 & 16.25$\pm$0.02 & -16.21 & &  \\
 &    &   & $I$ & 2454928.3$\pm$ 6.0 & 95.8$\pm$6.3 & 15.98$\pm$0.03 & -16.46 & &  \\
\hline
\end{tabular}
\begin{flushleft}
$^{a}$ The SN extinction is computed accounting for both Milky Way 
\protect\cite[][ see also Column 3 of Table \ref{Tab4}]{sch98} and host galaxy reddening components;
in the case of NOOS-005, since the host galaxy is projected behind LMC, we also accounted for the 
contribution of LMC. \\ $^{b}$ Computed using the B-band detections at days +429 and +432,
and assuming the same colour evolution as SN 1987A. Using the detection at day +216, we would obtain
M($^{56}$Ni) $\approx$ 0.35M$_\odot$.\\
$^{c}$ \protect\cite{ale08} and references therein. \\ $^{d}$ The broad delayed maximum 
typical of 1987A-like SNe is not visible in the B band; here we report the magnitude of the pseudo-plateau visible
after the early time B-band maximum.\\ $^{e}$ For a comprehensive information on the multi-band light curve parameters,
see Table 7 of \protect\cite{tad11}.\\

0 = This paper;
1 = Sandage $\&$ Tammann 1974;
2 = Yamagata $\&$ Iye 1982;
3 = Tsvetkov 1984; 
4 = Tsvetkov 1988; 
5 = Whitelock et al. 1989 (and references therein);
6 = Pastorello et al. 2005;
7 = Germany et al. 2004;
8 = Hamuy 2001;
9 = Hamuy $\&$ Pinto 2002;
10 = Kleiser et al. 2011;
11 = Taddia et al. 2011.
\end{flushleft}
\end{table*}

 Finally, adopting a mass of the compact remnant of about 2~M$_\odot$ (and assuming a negligible pre-SN mass loss), we
derive a final mass of 21~M$_\odot$ for the progenitor of SN 2009E, which is slightly higher than our
rough estimate obtained throught the O mass estimate (Section \ref{OI}). 

\section{SN 2009E and other 1987A-like events} \label{discussion}

Although 1987A-likeness was claimed for a number of objects in the past \citep[e.g. SNe 1923A, 1948B and 1965L,][]{sch88,you88,van89}, reasonable observational evidence exists for only a small
number of events \footnote{
Another type II SN, 2004ek, may show some similarity with this SN sub-group.  
Its light curve, published by \cite{tsv08}, is rather peculiar, showing a sharp early-time peak, mostly visible in the bluer bands, followed by a pseudo-plateau lasting a couple of months. However, especially in the redder bands, a sort of delayed broad peak is visible, analogous to the one characterizing SNe 1982F and 2000cb.
As major differences, SN 2004ek is significantly brighter ($M_V \approx$ -18) and its spectrum, showing an H$\alpha$ mostly in emission \citep{fil04} is rather peculiar. For this reason SN 2004ek has been rejected from our sample.}.
Additionally, for a few of them [SN 2004em, \cite{gal07}; SN 2005ci, \cite{arc09}] data are not published yet or are incomplete. Our limited 1987A-like sample includes
the objects that are briefly described in Appendix \ref{family} (and whose light curves are shown in Appendix \ref{appB}).
The sample has been selected among objects that are spectroscopically classified as type II (without showing type IIn-like spectral features), and with light curves showing a slow rise to maximum.

Main parameters of the host galaxies of our sample 
of 1987A-like events are listed in Table \ref{Tab4}. Remarkably, most SNe in our sample occured in late-type galaxies
(Scd or irregular). In addition, some of them were hosted by intrinsically faint (dwarf) galaxies,
which are thought to have low-metallicities \citep[see e.g.][]{you09}. However, at least half
of the objects of our sample occurred in luminous spirals. Although luminous spirals are believed to be rather
metal-rich, the 1987A-like events hosted in such galaxies occurred in peripheral -and hence more 
metal deficient- regions of the galaxies. Environments with low-to-moderate metallicity (slightly sub-solar) 
appear to be a fairly common characteristic for 1987A-like objects.

In Table \ref{Tab5} we report the values of some observational parameters for our SN sample. 
The magnitudes of the broad peaks seem to be correlated with the ejected $^{56}$Ni masses
(i.e. the objects with the brightest peak magnitudes are more $^{56}$Ni-rich).
The  $^{56}$Ni masses reported in Table \ref{Tab5} have been computed through a comparison between
the late-time {\sl uvoir} light curves of the different SNe with that of SN 1987A. The bolometric corrections
were computed with reference to SN 1987A. The derived $^{56}$Ni masses span a range between about 0.04 M$_\odot$
and 0.13 M$_\odot$\footnote{The lack of photometric points during the nebular phase did not allow us to obtain
reliable estimates for the $^{56}$Ni masses of SN 1982F.}, similar to
that observed in normal type IIP SNe, although cases of extremely $^{56}$Ni-poor 1987A-like events \citep[with $^{56}$Ni masses of the order of 10$^{-3}$M$_\odot$,][]{pasto04} have not been found so far.

Unfortunately, the sub-sample of 1987A-like SNe with an extensive data sets necessary for modelling is very small. 
This is necessary to constrain the properties of the progenitor stars.
The rather large ejected mass, the moderate amount of synthesized $^{56}$Ni, and the
small initial radius found for SN 2009E (Section \ref{model}) fit reasonably well within the framework of a relatively compact and massive progenitor, and are
in a good agreement with the results found in previous works on 1987A-like events \citep[see Table 8 in][]{tad11}. 
Using an older version of the semi-analytic modelling code than the one adopted 
here, \citet{pasto05}  obtained the following estimates for SNe 1987A and 1998A: E(1987A) = 1.6 foe, E(1998A) = 5.6 foe;  M$_{ej}$(1987A) = 18 M$_\odot$, M$_{ej}$(1998A) = 22 M$_\odot$; M$_{^{56}Ni}$(1987A) = 0.075 M$_\odot$, M$_{^{56}Ni}$ = 0.11 M$_\odot$;  R$_0$(1987A) $\gtrsim 5 \times 10^{12}$ cm, R$_0$(1998A) $\gtrsim 6 \times 10^{12}$ cm.
The major difference lies in the ejected $^{56}$Ni masses of SNe 1987A and 1998A, which are $\times$2 and $\times$3 
larger (respectively) than the one we have derived for SN 2009E (Section \ref{model}). 
Explosion parameters for a few additional 1987A-like SNe are also available in the literature \citep[see also ][]{tad11}. 
Using a one-dimensional Lagrangian radiation-hydrodynamics code, \citet{io11} derived the following parameters for the luminous 1987A-like SN 2000cb: E(2000cb) = 2 foe, M$_{ej}$(2000cb) = 17.5 M$_\odot$, M$_{^{56}Ni}$(2000cb) = 0.10 M$_\odot$ and R$_0$(2000cb) $= 3 \times 10^{12}$ cm. These values  are slightly different from those derived by  \citet{utr11} by modelling the spectroscopic and photometric data of SN 2000cb with a different hydrodynamic code. They inferred a presupernova radius of $2.4 \times 10^{12}$ cm, an ejected mass of 22.3 M$_\odot$, an energy of 4.4 foe, and a mass of radioactive $^{56}$Ni of 0.083 M$_\odot$\footnote{Note, however, that the distance moduli adopted  by \cite{pasto05} for the host of SN 2009E ($\mu$ = 32.41) and by  \cite{io11} and \cite{utr11} for the host of SN 2000cb ($\mu = 32.39$) are different from those adopted here and reported in Table \ref{Tab4}.}. With a semi-analytic approach, \citet{tad11} estimated relevant physical parameters for two 1987A-like events, SNe 2006V and 2006au: E(2006V) = 2.4 foe, E(2006au) = 3.2 foe; M$_{ej}$(2006V) = 17.0 M$_\odot$,  M$_{ej}$(2006au) = 19.3 M$_\odot$; M$_{^{56}Ni}$(2006V) = 0.127 M$_\odot$, M$_{^{56}Ni}$(2006au) $\leq$ 0.073 M$_\odot$; R$_0$(2006V) $= 5.2 \times 10^{12}$ cm, R$_0$(2006au) $= 6.3 \times 10^{12}$ cm.

It is evident that 1987A-like SN explosions have similar kinetic energies (a few $\times$ 10$^{51}$ erg) and ejected masses (17 to 22 M$_\odot$), whilst they appear to span a factor of 3 in $^{56}$Ni masses \citep[see also Table 8 in][ for a summary of the parameters of the best-studied 1987A-like events]{tad11}.
It is intriguing to note that the modelling reported above suggest that {\sl all the 1987A-like SNe would be produced by the explosions of BSG progenitors that have initial radii ranging between 35 and 90 R$_\odot$, and with final masses around 20 M$_\odot$}. These progenitor masses are systematically higher than those estimated for the RSG progenitors of classical SNe IIP \citep[e.g.][]{sma09}.

\section{Rate of 1987A-like events} \label{rate}

While it would be of interest to know the relative rate of 1987A-like events vs. normal type II SNe, the
heterogeneous properties of this class and the uncertainties in the sample population from which 
the events reported in Table \ref{Tab4} are extracted prevent us from providing any accurate estimates.
With the aim to derive an educated guess, the completeness of the SN searches with
respect to the 1987A-like objects list can be assessed by comparing the distance modulus
distribution of 1987A-like events with that of a suitable reference sample. For the
latter, we extracted from the Asiago Supernova Catalogue the list of all type II SNe.
If we limit the distance modulus to about 35 mag (at higher redshift incompleteness appears to be significant)  
it seems that 1987A-like events would account for less than 1.5$\%$ of all type II SNe.
This value, that can be considered as a lower limit for the relative rate of 1987A-like events, 
is very similar to that derived by \cite{io11} based on the two 1987A-like SNe discovered by the LOSS.

However, the identification of 1987A-like events demands a fairly good light curve
coverage that is not available for many of the type II SNe listed in the catalogue.
Actually, reasonably good light curves of type II (IIP, IIL, IIb, IIn) SNe exist 
for about 300 objects. This approximate estimate was obtained including both SNe whose data 
have been published in literature, and SNe spectroscopically classified whose photometric 
follow-up information is available in public pages of current SN monitoring programs\footnote{For example, the ESO/TNG 
large program on Supernova Variety and Nucleosynthesis Yields
({\sl http://graspa.oapd.inaf.it/}), the Caltech Core-Collapse Program 
({\sl http://www.astro.caltech.edu/ $^\sim$avishay/cccp.html}) or the Carnegie Supernova Project
({\sl http://csp1.lco.cl/ $\sim$cspuser1/PUB/CSP.html})}.
From this estimate, we infer that 1987A-like events comprise about 4$\%$
of all type II SNe. This has to be considered as a sort of upper limit, since we expect
that the astronomical community has been investing more efforts to follow peculiar 1987A-like 
SNe rather than other type II SN sub-types because of their intrinsical rarity.
Therefore, we estimate that SNe photometrically similar to SN 1987A comprise  between $\sim$1.4 and 4$\%$
of all type II SNe, corresponding to about 1-3$\%$ of all core-collapse SNe in a volume-limited
sample \citep[e.g.][]{sma09,li11}.

\section{Summary} \label{summary}

New data for the peculiar type II SN 2009E have been presented. Its light curve is very similar to that of SN 1987A, with a few main differences: the light curve peak is shifted in phase by about +10 days with respect to that of SN 1987A, its luminosity
is fainter than that of SN 1987A and the amount of $^{56}$Ni deduced from the luminosity of the radioactive tail is
about 0.04M$_\odot$. This is the lowest ejected $^{56}$Ni mass estimated for any 1987A-like event studied so far. The spectra of SN 2009E appear to be 
 different from those of SN 1987A (and similar events), while they share major similarities with those of sub-luminous, $^{56}$Ni-poor SNe IIP (e.g. SN 2005cs) in terms of line velocities and relative line strengths.

In analogy with what has been deduced for SN 1987A, also the observed properties of SN 2009E are consistent with those expected from the explosion of a BSG star.
Modelling the data of SN 2009E, we obtained a relatively under-energetic explosion ($E = 0.6$ foe), an initial radius of 
of $7 \times 10^{12}$ cm and an ejected mass of $19 M_\odot$, pointing toward a relatively high-mass
precursor star ($21 M_\odot$), very similar to the BSG progenitor inferred for SN 1987A.

Finally, we compared the observed properties of a dozen of 1987A-like SNe available in the literature, and found that this subgroup of type II SNe shows significant variety in the explosion parameters and in the characteristics of the host galaxies (although they seem to prefer moderately metal-poor environments). This would suggest for these objects a distribution in the space of parameters which is similar to that observed in more classical type IIP SNe, although they are intrinsically rare ($\lesssim$ 3$\%$ of all core-collapse SNe in a volume-limited sample). 

\begin{acknowledgements}
This work is based in part on observations collected at the Italian 3.58-m Telescopio Nazionale Galileo (TNG), the 2.56-m Nordic Optical Telescope, the 2.2-m Telescope of the Centro Astron\'omico Hispano Alem\`an (Calar Alto, Spain) and  the 1.82-m Copernico Telescope on Cima Ekar (Asiago, Italy). The TNG is operated by the Fundaci\'on Galileo Galilei of the Instituto Nazionale di Astrofisica at the Spanish Observatorio del Roque de los Muchachos of the Instituto de Astrofisica de Canarias. We thank the support astronomers at the TNG, the 2.2-m Telescope at Calar Alto and the Nordic Optical Telescope for performing the follow-up observations of SN 2009E.
SB, FB, EC, MT and AH are partially supported by the PRIN-INAF 2009 with the project `Supernovae Variety and Nucleosynthesis Yields'. MLP acknowledges financial support from the Bonino-Pulejo Foundation. SM acknowledges financial support from the Academy of Finland
(project: 8120503).
We are grateful to the TriGrid VL project and the INAF-Astronomical Observatory of Padua for the use of 
computer facilities. 

EP acknowledges the project Skylive of the Unione Astrofili Italiani (UAI) and the 
Skylive Remote Facility - Osservatorio B40 Skylive, Catania, Italy ({\it http://www.skylive.it}).
 We thank the amateur astronomers of the Associazione Astronomica Isaac Newton ({\it http://www.isaacnewton.it}) 
and of the Associazione Astronomica Cortina ({\it http://www.cortinastelle.it/}) for providing images of 
SN 2009E obtained during the SMMSS and CROSS supernova search programs: 
G. Iacopini (Osservatorio di Tavolaia, S. Maria a Monte, 
Pistoia), R. Mancini and F. Briganti (Stazione di Osservazione di Gavena, Cerreto Guidi, Italy), A. Dimai 
(Osservatorio di Col Driusc\'e, Cortina, Italy). We also thank 
J.M. Llapasset, O. Trondal, Boyd, for sharing their images with us, I. Bano and A. Englaro
(Osservatorio Astronomico Polse di Cougnes) for their help with observations,
A. Siviero for allowing ToO observations of SN 2009E at the 1.82-m Telescope of Asiago (Mt. Ekar, Italy)
and T. Iijima for providing us the B$\&$C spectrum obtained at the 1.22-m Galileo Telescope 
(Osservatorio Astrofisico di Asiago). AP is grateful to Mario Hamuy, Io Kleiser, Dovi Poznanski and 
Lisa Germany for sharing their data of SNe 2000cb and 1998bt.

This manuscript made wide use of information contained in the Bright Supernova web pages 
(maintained by D. Bishop), as part of the Rochester Academy of Sciences 
({\it http://www.RochesterAstronomy.org/snimages/}). We also used information collected 
from the the web site SNAude des Supernovae ({\it http://www.astrosurf.com/snaude/}).
The paper is also based in part on data collected at the Akeno 50 cm Telescope (Akeno Observatory/ICRR, Yamanashi, Japan) 
and obtained from the SMOKA, which is operated by the Astronomy Data Center, National Astronomical Observatory of Japan;
and at the Faulkes North Telescope and obtained through the Faulkes Telescopes Data Archive.

\end{acknowledgements}

\bibliographystyle{aa}

\begin{appendix}

\section{Basic information on the sample of SN 1987-like transients} \label{family}

\begin{enumerate}

\item {\bf SN 1987A} is the prototype of this family of H-rich core-collapse SNe, and -since it exploded in the nearby Large Magellanic
Cloud- it is the best studied SN ever. 
The SN was discovered on 1987 February 24th by Shelton. Strong constraints on the time of the core-collapse
came from the detection of a neutrino burst on February 23.316 UT by IMB and Kamiokande II \citep{bio87,hir87}.
Extensive data sets at all wavelengths are provided by a large number of publications 
\citep[e.g.][]{men87,cat87,cat88,cat89,whi88,whi89,ham88,phi88,phi90,pun95}. More recent papers have unveiled
the complex structure of the circumstellar environment of SN 1987A and its interaction with the SN ejecta 
\citep[e.g.][]{bor97}. SN 1987A is one of the few objects for which we have robust contraints on the nature of the progenitor 
star. Pre-explosion images of the source at the SN position (Sk $-$69 202) showed it to be a blue (B3 I) supergiant 
\citep{gil87,son87} whose mass is estimated to be around 20 M$_\odot$ \citep[see][ for a review]{arn89}. 
A small initial radius of the progenitor star is also used to explain the unusual, slow rise to maximum
of the light curves of SN 1987A. In this paper we adopt as distance modulus
of LMC and as total reddening toward SN 1987A the values of $\mu$ = 18.50 mag \citep{sak04} and $E(B-V)$ = 0.19
mag \citep{arn89}.

\item {\bf SN 1909A} is a historical SN that exploded in M101. Its photometric evolution shows striking similarity with 
that of SN 1987A \citep[][ see also Appendix \ref{appB}]{you88,pat94}. 
A well-sampled light curve from photographic plates rescaled to the B-band system was published by \cite{san74}.
We assume negligible extinction toward SN 1909A and a distance modulus of $\mu$ = 29.34 mag \citep[][ see Table \ref{Tab4} for details]{riz07}. Although multiband observations are missing for this event, the B-band absolute magnitude at maximum ($M_B \approx -$15.9, see 
Table \ref{Tab5}) suggests that SN 1909A is one of the intrinsically brightest SNe in our sample. 

\item {\bf SN 1982F} in NGC 4490 is a poorly followed under-luminous type II SN.
Sparse data around maximum from photographic plates were published by \cite{yam82} and \cite{tsv84,tsv88}. The light curves share 
remarkable similarity with SN 2000cb (see below), more than with SN 1987A. No spectrum exists for this object to our knowledge.

\item {\bf SN 1998A} is a well studied 1987A-like event which exploded in a spiral arm of the SBc galaxy
IC 2627. The SN was discovered on 1998 January 6 by the Automated Supernova Search Program of the Perth
Astronomy Research Group \citep[PARG,][]{wil98}. 
Optical photometry and spectroscopy were published by \cite{woo98} and \cite{pasto05}. Adopting the HyperLeda recessional velocity
corrected for local infall onto Virgo v$_{Vir}$ = 1976 km s$^{-1}$, we obtain a 
distance of 27.4 Mpc ($\mu$ = 32.19 mag). Since there was no evidence of additional extinction in the host galaxy,
we  adopt the same total reddening estimate as in \cite{pasto05}, i.e. $E(B-V)$ = 0.12 mag.

\item {\bf SN 1998bt} was discovered in the Abell cluster 1736 by the Mount Stromlo Abell cluster supernova search team
on March 10, 1998 \citep{ger98}. Initially, no background galaxy was seen to be associated with the SN. However,
subsequent deep imaging of the SN field revealed a very faint parent galaxy of $R$ = 23.4 \citep{ger04}.
Unfortunately, spectroscopic classification for this SN does not exist. However, the overall behaviour of the light curve
(see Figure \ref{FigA1}) is reminiscent of that of SN 1987A. 

\item {\bf SN 2000cb} in IC 1158 is one of the best followed 1987A-like events in our sample. Discovered on April 27.4 2000 
using the 0.8-m Katzman Automatic Imaging Telescope \citep[KAIT,][]{pap00}, it was classified as a young type II 
SN by \cite{jha00} and \cite{ald00}. Optical photometric and spectroscopic observations have been presented by 
\cite{ham01} and \cite{io11}. The photometric evolution of SN 2000cb is 
different from that of SN 1987A.
The light curve in the $B$ band shows a maximum at $\approx$ 40 days 
past core-collapse, followed by a slow decline up to 60 days and a short pseudo-plateau. At about 90 days a steep
post plateau decline to the radioactive tail is visible, as in normal type IIP SNe.
The evolution in the $VRI$ bands is somewhat different, showing  broad light curve peaks at 60$-$80 days (depending on the band) 
after core-collapse \citep{ham01,io11}. The light curve peaks are significantly broader than those
of SN 1987A. We adopted the observed parameters of SN 2000cb as presented in \cite{ham01}. 
The explosion epoch, the adopted distance and the total reddening
are those of \cite{ham01}. Their values are listed in Tables \ref{Tab4} and \ref{Tab5}.

\item {\bf OGLE-2003-NOOS-005} never had a SN designation. It was discovered by the 
Optical Gravitational Lensing Experiment \citep[OGLE,][]{uda03} collaboration at $\alpha = 5^h55^m38^s.07$ and $\delta = -68^\circ55'47\farcs1$ (J2000.0).
It exploded in a faint spiral galaxy labelled by the NED database  
as 2MASX J05553978-6855381, behind LMC, and falls in the OGLE field LMC198.6 9 \citep{uda08}.
The well-followed I-band light curve of OGLE-2003-NOOS-005\footnote{available via ftp through
the OGLE-III (2003 season) web site {\it http://ogle.astrouw.edu.pl/ogle3/ews/NOOS/2003/noos.html}} well matches 
that of SN 1987A. Unfortunately multiband observations  for this SN do not exist.
The redshift of the host galaxy was spectroscopically determined measuring the positions of a few selected spectral
features of a host galaxy spectrum obtained on 2009 August 14 with NTT (equipped
with EFOSC2 and grism 11; resolution = 22 \AA). Despite the low signal-to-noise, we identified a few absorption features (Ca II H$\&$K, the $g$-band $\lambda$4200, Mg Ib $\lambda$5173), and H$\alpha$ in emission. 
In addition the emission line of [O II] $\lambda$3727 was marginally detected. 
This allowed to fix the redshift at  z = 0.0302 $\pm$ 0.0008
(v$_{rec} \approx$ 9151 $\pm$ 236 km s$^{-1}$), which corresponds to a distance of about 127 Mpc ($\mu$ = 35.52 mag). 
These values are remarkably similar to those reported by NED \citep[v$_{rec} \approx$ 9176 $\pm$ 45 km s$^{-1}$;][]{jon09}.
and HyperLeda (v$_{rec} \approx$ 9177 $\pm$ 60 km s$^{-1}$). The recessional velocity corrected for Local Group infall
into the Virgo Cluster is slightly lower (v$_{Vir} \approx$ 8882 km $s^{-1}$). Adopting 
the Hubble Constant value of H$_0$ = 72 km s$^{-1}$ Mpc$^{-1}$, we obtain a distance of 123.4 Mpc, i.e.
distance modulus $\mu$ = 35.46 mag. 
We also assume in our analysis a total reddening toward the transient of $E(B-V)$ = 0.16 $\pm$ 0.10 mag. 
This was determined including the contribution of the Galaxy  $E_{MW}(B-V)$ = 0.075 mag \citep{sch98} and
the reddening at the position of 2MASX J05553978-6855381 due to the intrinsic contribution of LMC that was
computed using the maps of \cite{zar04} ($E_{LMC}(B-V)$ = 0.087 mag).
The I-band light curve of this SN is shown in Appendix B.

\item {\bf SN 2004em} was discovered by \cite{arm04} on Sept. 14, 2004 when the SN was at $mag$ = 17.5.
The transient, later classified as a young type II by \cite{fil04}, was hosted in an Sc galaxy (IC 1303). 
The object was significantly reddened by Galactic dust \citep[$E(B-V)$ = 0.108, from][]{sch98}. Preliminary information
on the Caltech Core-Collapse Project light curve has been provided by \cite{gal07}.


\item {\bf SN 2005ci}, a LOSS discovery \citep{mad05}, was classified as a type II SN by \cite{mod05} on the basis
of the presence of P-Cygni Balmer lines. Its peculiar light curve was first noted by \cite{arc09}, and some early-time unfiltered photometry
was provided by \cite{io11}, showing a clear light curve rise by about 2.5 mag in $\sim$ 45 days. Unfortunately, with the modest amount of data
available so far, we cannot estimate the main SN parameters. The host galaxy, NGC 5682, is an Sb-type; the Galaxy reddening is modest,
$E(B-V)$ = 0.033 mag \citep{sch98}, but the relatively red continuum shown by the spectrum presented by \cite{io11} (see their Figure 12) suggests a
non negligible host galaxy reddening. The narrow host galaxy Na ID feature is not clearly detected and the amount of internal reddening cannot 
be constrained.

\item {\bf SN 2006V} was discovered in the course of the Taiwan Supernova Survey on
Feb. 4, 2006 UT at a magnitude of about 18 \citep{chen06}. The object was classified as a type II SN after maximum by \cite{blo06} and extensively followed by the Carnegie Supernova Project collaboration, who noted the slowly rising, 1987A-like light curve \citep{tad11}. The galaxy hosting SN 2006V is 
UGC 6510, which is classified as a face-on spiral (SABc, HyperLeda source) suffering of small Galactic reddening \cite[$E(B-V)$ = 0.029 mag,][]{sch98}. 
The lack of detection of narrow Na ID absorption at the redshift of UGC 6510, suggest no additional host galaxy extinction toward SN 2006V \citep{tad11}.
A comparison between the multi-band light curves of SNe 2006V and 1987A can be found in Figure 3 of \citet{tad11}. 

\item {\bf SN 2006au} was a discovery of the Tenagra Observatory Supernova Search using the 0.35-m Tenagra
telescope in Oslo (on Mar. 7.20 UT). At the discovery, the object had $mag$ 17.2 \citep{tro06}. SN 2006au was later classified by the Nearby Supernova Factory as a type II SN \citep{bla06}. Again, a comprehensive study is presented in \citet{tad11} and comparisons with the light curves of SN 1987A are shown in their Figure 4. 
The host galaxy, UGC 11057, is a late spiral (possibly an Sc-type, according to HyperLeda) with rather large Galactic reddening, i.e. $E(B-V)$ = 0.172 mag \citep{sch98}.
However, \citet{tad11} found that SN 2006au suffered also of large extinction due to dust in the host galaxy, deriving a total reddening of $E(B-V)_{tot}$ = 0.312 mag. 
\end{enumerate}

\section{Light Curves of 1987A-like events} \label{appB}

   \begin{figure*}
   \centering
   {\includegraphics[angle=0,width=9.1cm]{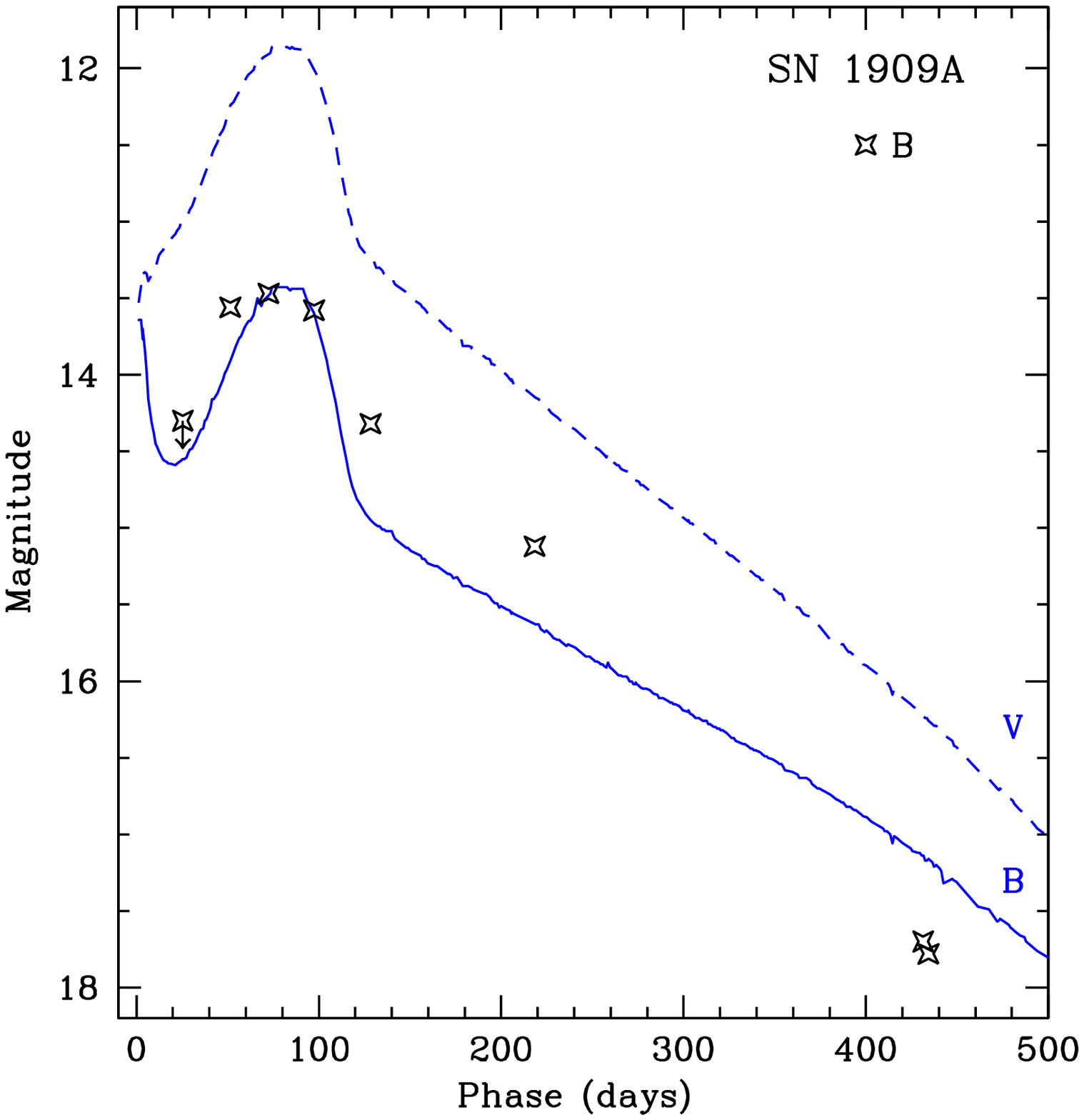}
   \includegraphics[angle=0,width=9.1cm]{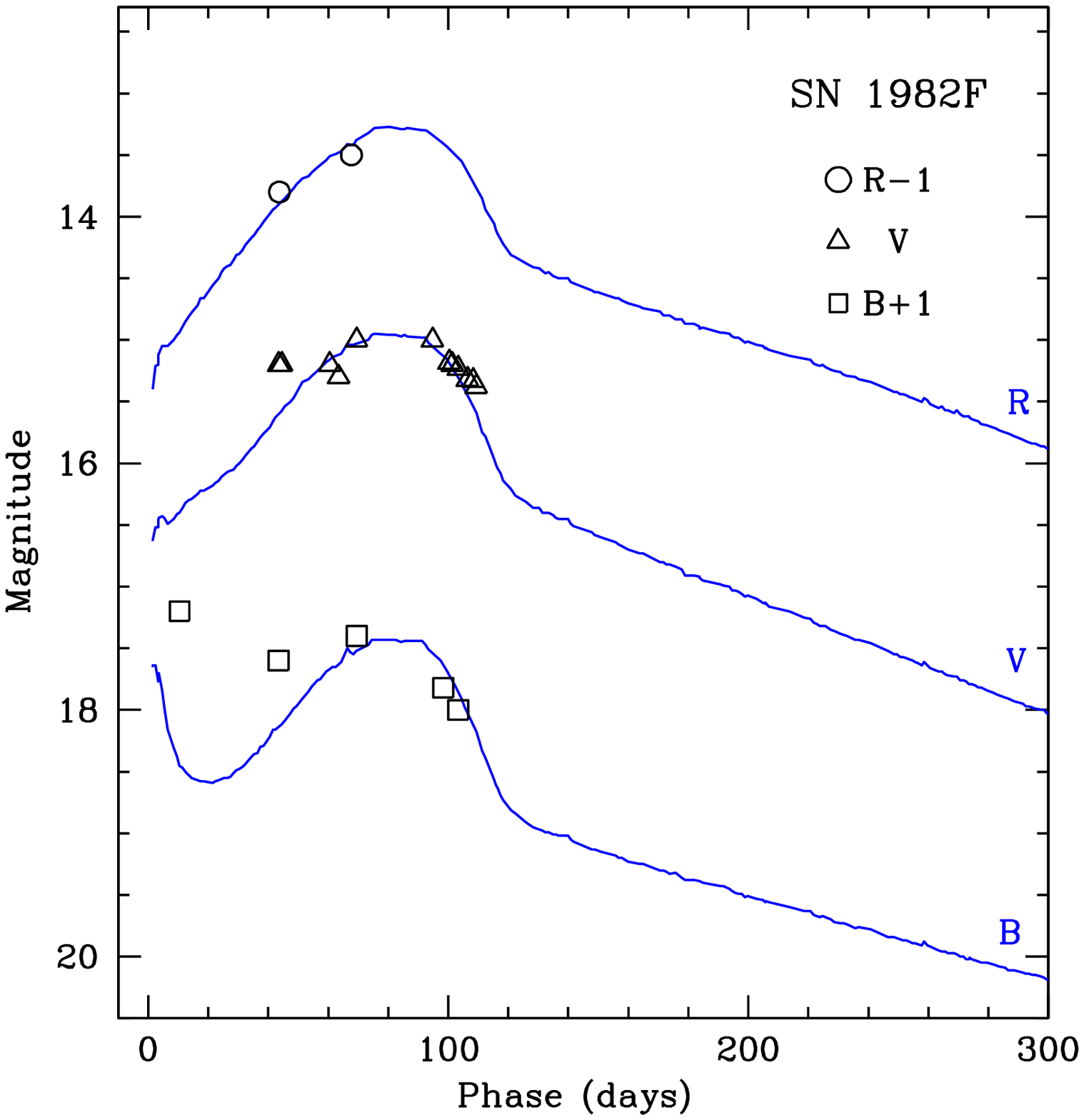}
   \includegraphics[angle=0,width=9.1cm]{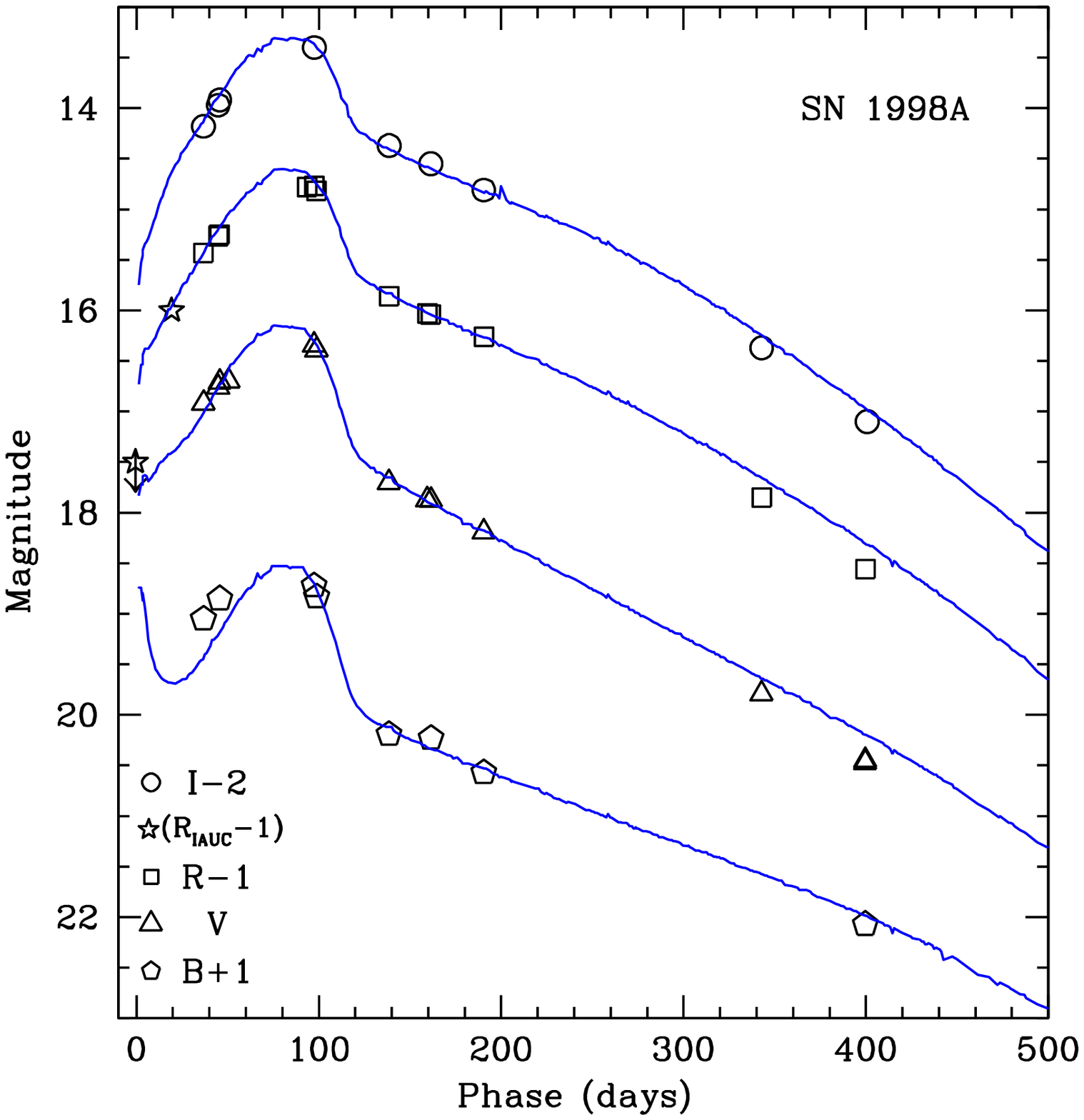}
   \includegraphics[angle=0,width=9.1cm]{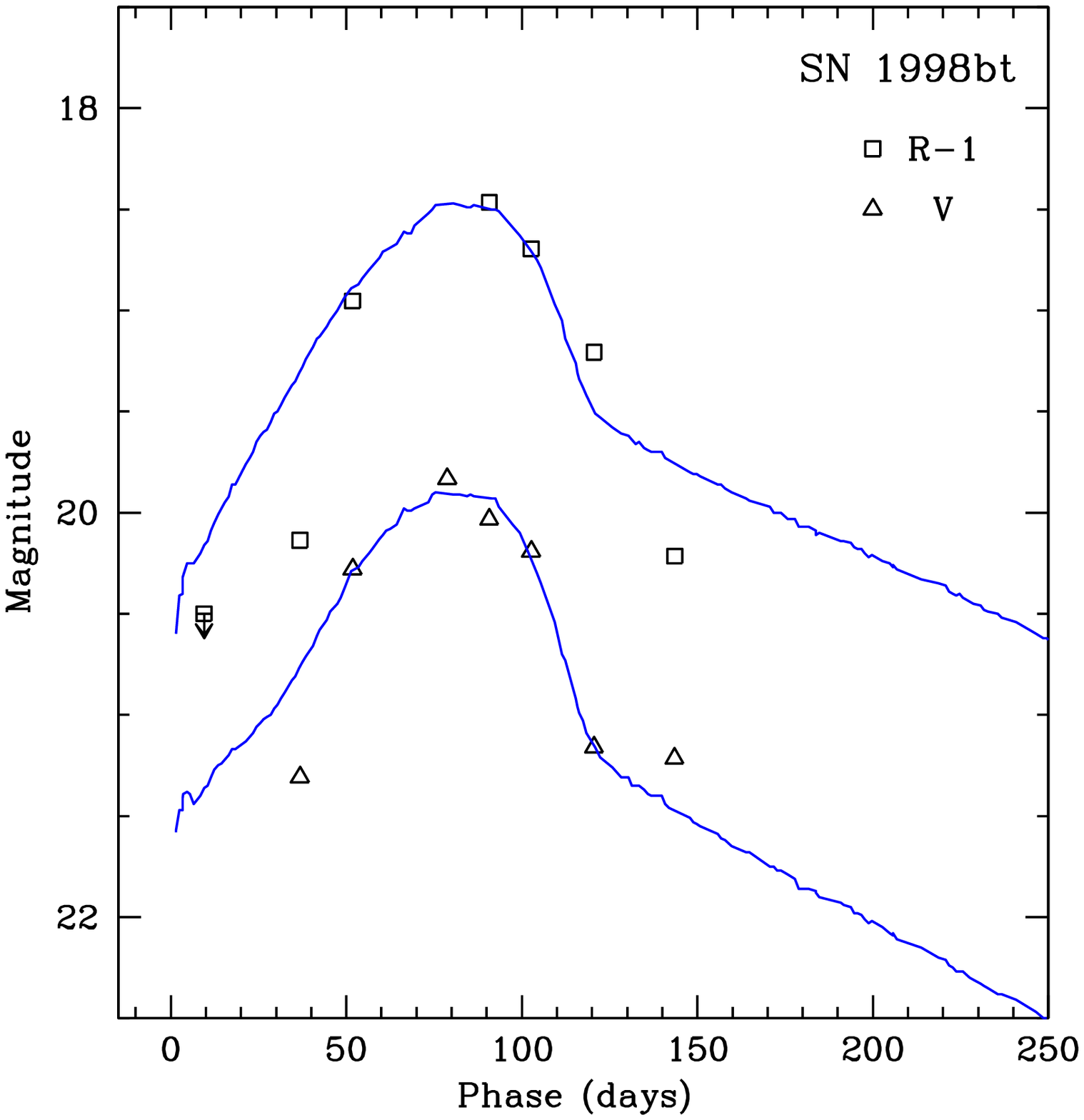}}
   \caption{Light curves of 1987A-like events: first group of 4 events. 
The solid (or dashed) blue lines represent the light
curves of SN 1987A, our template. The light curves of SN 1987A have been 
arbitrarily shifted in magnitude to match the peaks of the other transients.}
              \label{FigA1}
    \end{figure*}

\begin{figure*}
   \centering
   {\includegraphics[angle=0,width=9.1cm]{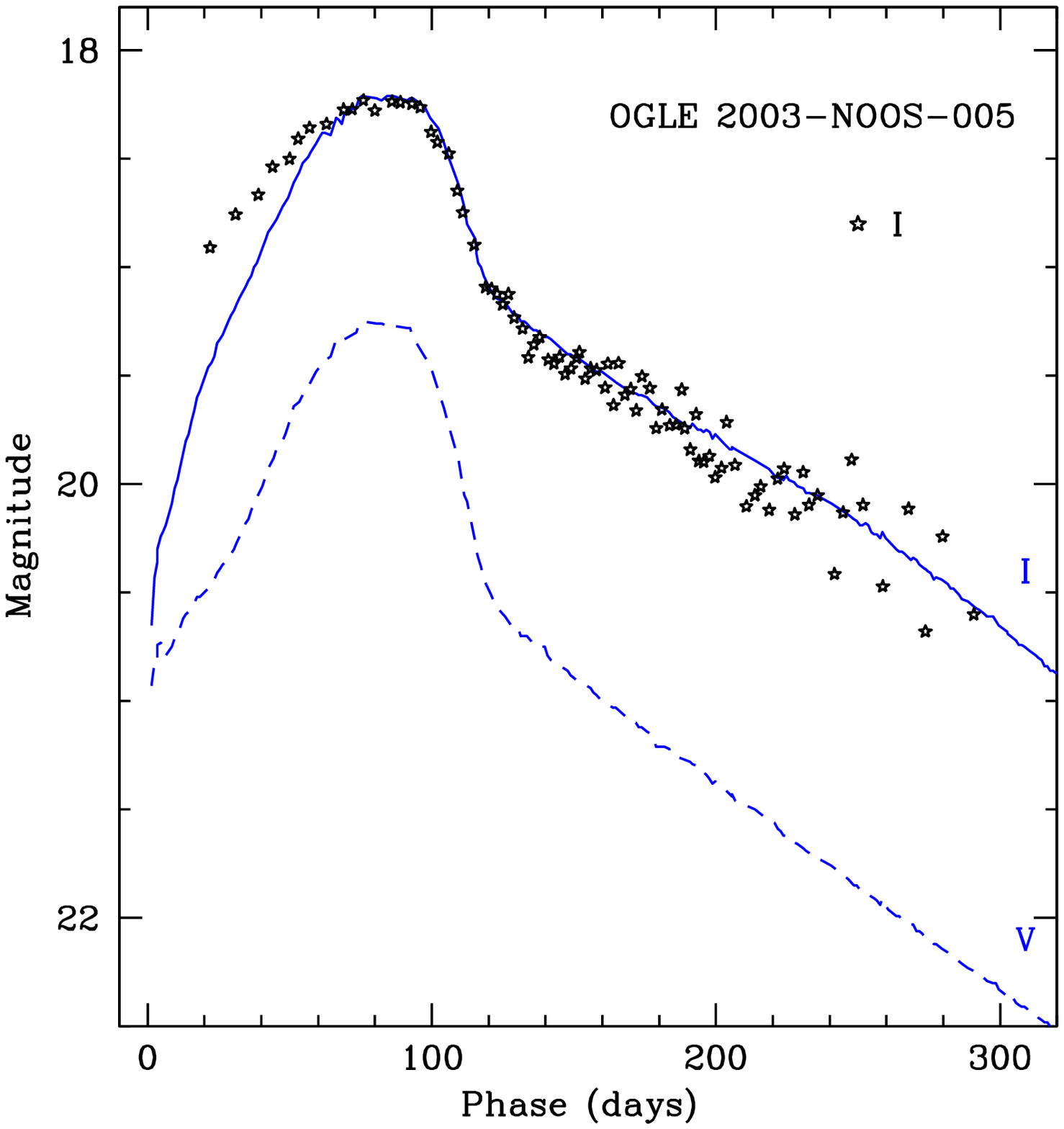}
   \includegraphics[angle=0,width=9.1cm]{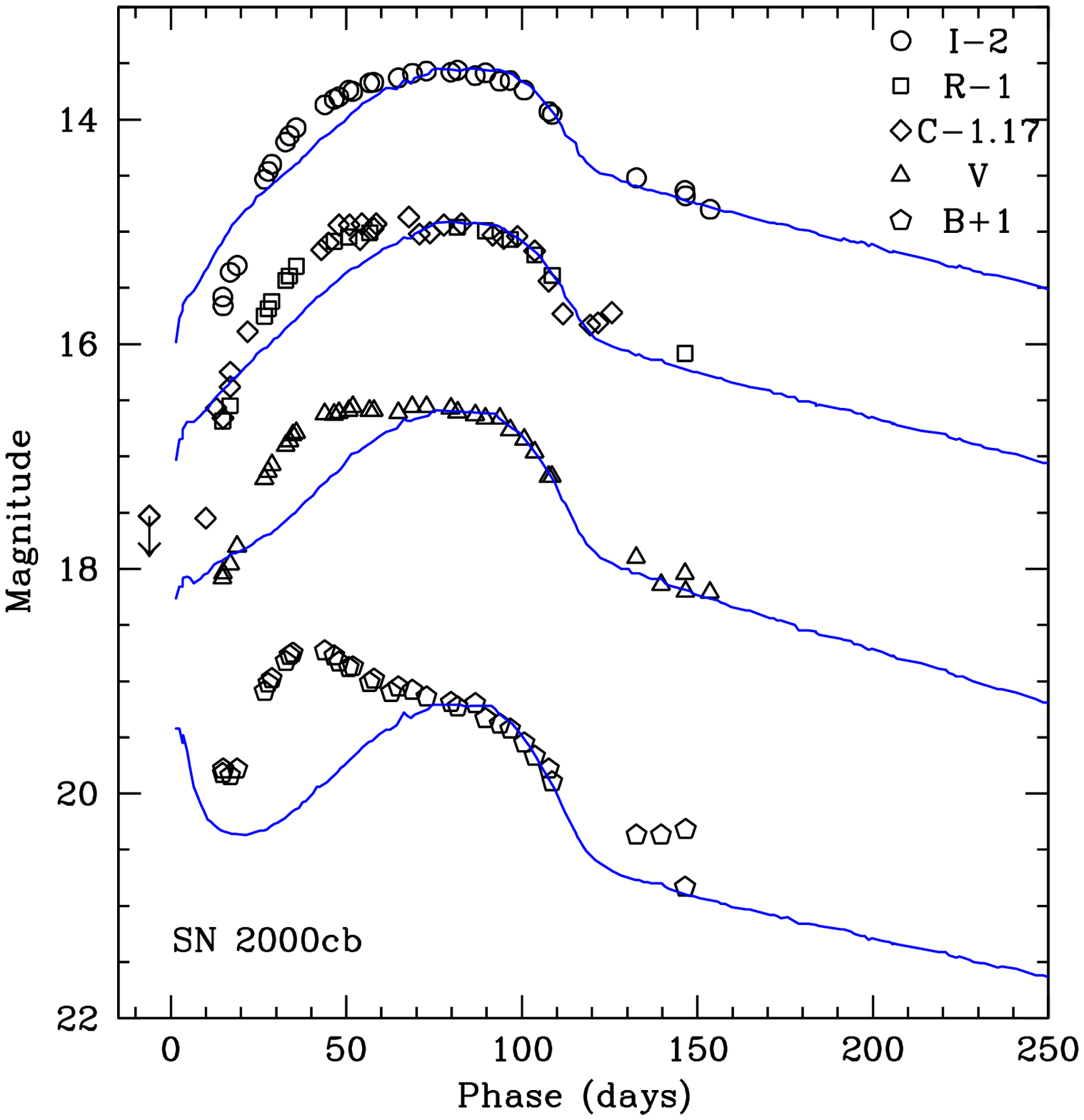}
   \includegraphics[angle=0,width=9.1cm]{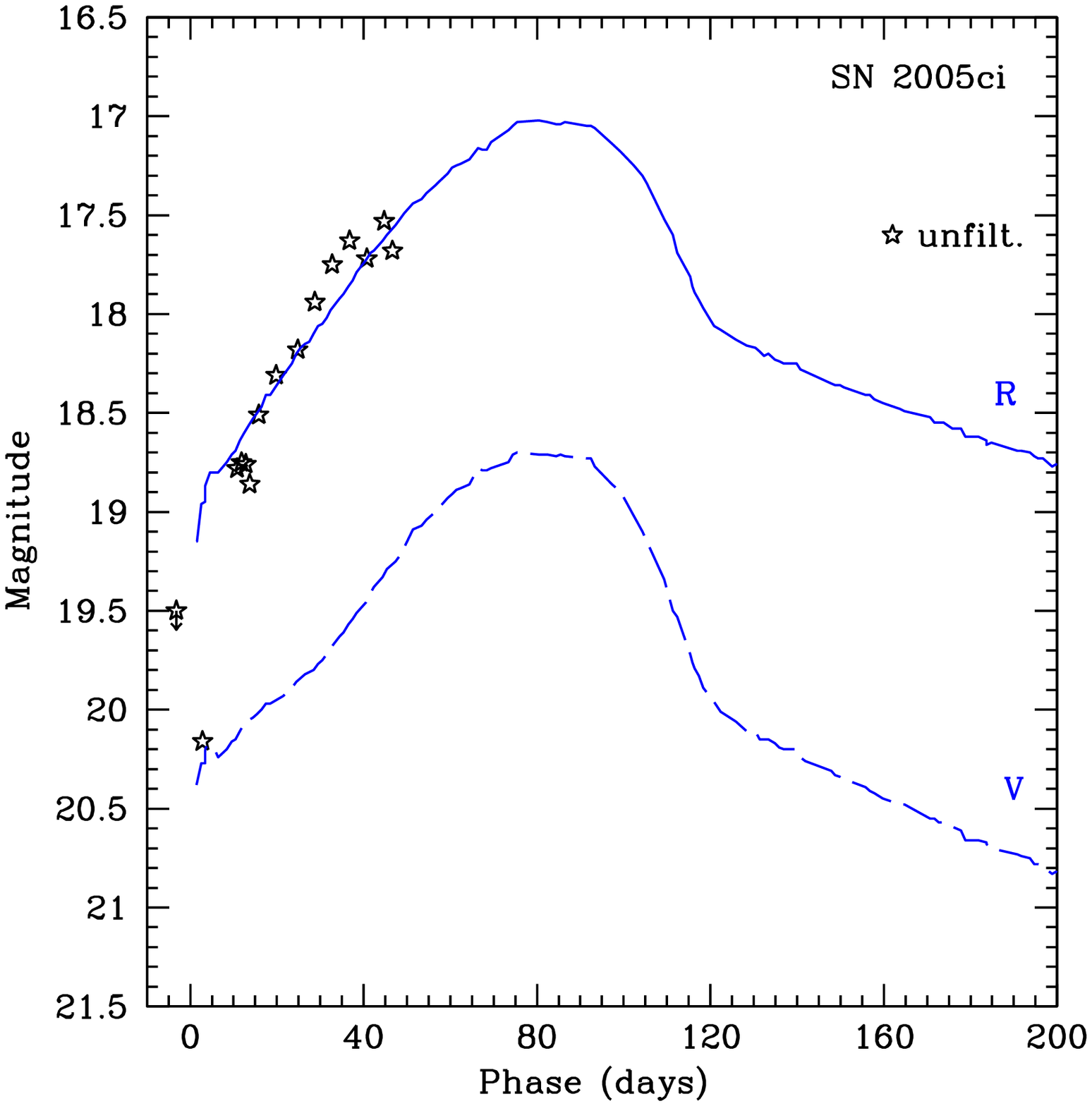}}	
   \caption{Light curves of 1987A-like events: second group of 3 events. 
The solid (and dashed) blue lines represent the light
curves of SN 1987A, our template. The light curves of SN 1987A have been arbitrarily shifted in magnitude 
to match the peaks of the other transients.}
              \label{FigA2}
    \end{figure*}

In this section, the light curves of eight 1987A-like type II SNe  are compared with those of the prototype 
SN 1987A (solid or dashed blue lines, see Figures \ref{FigA1} and \ref{FigA2}). 
The light curves of SN 1987A are arbitrarily shifted in magnitude to match the peak magnitudes
of the other objects. In the comparison, we assumed the $JD$s listed in Table \ref{Tab4} as explosion epochs.
For a few individual events (e.g. SNe 1982F or 2000cb) the 1987A-likeness is more evident 
in the red bands than in the blue ones. As mentioned in Sect. \ref{discussion}, data for SN 2004em have not been published yet.

\end{appendix}

\end{document}